\documentclass[
	arxiv
	]{medphyspaper}

\usepackage[utf8]{inputenc}
\usepackage[T1]{fontenc}

\usepackage{todonotes}
\usepackage{mathtools}
\usepackage{mathrsfs}

\usepackage{acronym}

\usepackage{pdftexcmds}

\usepackage{xpatch}

\ifarxivpreprint
	\usepackage[final]{changes}
\else
	\usepackage[draft]{changes}
\fi

\newacro{APM}{Analytical Probabilistic Modeling}
\newacro{IMRT}{intensity-modulated radiation therapy}
\newacro{DVH}{dose-volume histogram}
\newacro{CDF}{cumulative distribution function}
\newacro{PDF}{probability density function}
\newacro{VOI}{volume of interest}
\newacro{DVCM}{dose-volume coverage map}
\newacro{CTV}{clinical target volume}
\newacro{PTV}{planning target volughme}
\newacro{GTV}{gross tumor volume}
\newacro{OAR}{organ at risk}
\newacro{QI}{plan quality indicator}
\newacro{EUD}{equivalent uniform dose}

\usepackage{apmstyle}

\usepackage{hyperref}
\usepackage[capitalize,noabbrev]{cleveref}

\usepackage{xspace}
\usepackage{xparse}
\usepackage{csquotes}

\usepackage{siunitx}
\usepackage{booktabs}

\usetikzlibrary{external}
\usetikzlibrary{positioning,shapes,shadows,arrows.meta,arrows,backgrounds}

\pgfplotsset{compat=newest}

\newif\iftikzforceexternal

\tikzforceexternaltrue

\NewDocumentCommand{\includetikz}{O{.tikz} O{tikz/} O{tikz_ext/} m}{%
	\iftikzforceexternal%
		\includegraphics{#3#4.pdf}%
	\else
		\tikzsetnextfilename{#3#4}%
		\input{#2#4#1}%
	\fi
}
\makeatletter
\renewcommand{\todo}[2][]{\tikzexternaldisable\@todo[#1]{#2}\tikzexternalenable}
\makeatother

\renewcommand{\vecbf}[1]{\boldsymbol{#1}}
\newcommand{\numbix}{\ensuremath{B}}
\newcommand{\numvox}{\ensuremath{V}}

\makeatletter
\newcommand{\ie}{i.\,e.,\@\xspace}
\newcommand{\eg}{e.\,g.\@\xspace}
\makeatother

\renewcommand{\transposed}{T}

\definecolor{gre1}{rgb}{0.00000,0.47170,0.46040}%
\definecolor{dre}{rgb}{0.49060,0.00000,0.00000}%
\definecolor{ora}{rgb}{1.00000,0.60000,0.20000}%
\definecolor{blu}{rgb}{0.00000,0.00000,0.50900}%
\definecolor{gra}{rgb}{0.50000,0.50000,0.50000}%
\definecolor{graDark}{rgb}{0.20000,0.20000,0.20000}%
\definecolor{red}{rgb}{0.73725,0.07059,0.23922}%
\definecolor{gre}{rgb}{0.00392,0.49412,0.20000}%
\definecolor{dkfzlB}{rgb}{0.66275,0.73725,0.83529}%
\definecolor{dkfzmB}{rgb}{0.27451,0.46275,0.67843}%
\definecolor{dkfzdB}{rgb}{0.00000,0.29412,0.55686}%
\definecolor{lightmpg}{rgb}{0.50000,0.73585,0.73020}%
\definecolor{lightdre}{rgb}{0.74530,0.50000,0.50000}%
\definecolor{lightblu}{rgb}{0.50000,0.50000,0.75450}%
\definecolor{lightora}{rgb}{1.00000,0.80000,0.60000}%
\definecolor{lightgre}{rgb}{0.38431,0.79216,0.01569}%
\definecolor{lightgre2}{rgb}{0.77647,0.98039,0.56078}%
\definecolor{lightgre3}{rgb}{0.82353,1.00000,0.66275}%
\definecolor{darkgre}{rgb}{0.14510,0.43529,0.07451}%
\definecolor{darkora}{rgb}{0.76863,0.53725,0.00000}%
\definecolor{orared}{rgb}{0.85490,0.34902,0.12941}%
\definecolor{plotblue}{rgb}{0.45490,0.67840,0.81960}%
\definecolor{plotorange}{rgb}{0.84310,0.18820,0.15290}%
\definecolor{plotyellow}{rgb}{0.99220,0.68240,0.38040}%
\definecolor{plotblue_light}{rgb}{0.72745,0.83920,0.90980}%
\definecolor{plotorange_light}{rgb}{0.92155,0.59410,0.57645}%
\definecolor{plotyellow_light}{rgb}{0.99610,0.84120,0.69020}%
\definecolor{plotblue_grayed}{rgb}{0.72842,0.78430,0.81960}%
\definecolor{plotorange_grayed}{rgb}{0.84310,0.67937,0.67055}%
\definecolor{plotyellow_grayed}{rgb}{0.99220,0.91475,0.83925}%
\definecolor{plotblue2}{rgb}{0.27060,0.45880,0.70590}%
\definecolor{plotorange2}{rgb}{0.95690,0.42750,0.26270}%
\definecolor{plotyellow2}{rgb}{0.45490,0.67840,0.81960}%
\definecolor{plotblue2_light}{rgb}{0.63530,0.72940,0.85295}%
\definecolor{plotorange2_light}{rgb}{0.97845,0.71375,0.63135}%
\definecolor{plotyellow2_light}{rgb}{0.72745,0.83920,0.90980}%

\newlength\figurewidth
\newlength\figureheight
\newlength\subfigwidth

\colorlet{myblue}{plotblue}
\colorlet{myred}{plotorange}
\colorlet{myyellow}{plotyellow}
\colorlet{mygreen}{darkgre}

\title{Analytical probabilistic modeling of dose-volume histograms}
\runningtitle{Analytical probabilistic modeling of DVHs}

\author[DKFZ,HIRO]{Niklas Wahl}

\address[DKFZ]{German Cancer Research Center -- DKFZ, Im Neuenheimer Feld 280, 69120 Heidelberg, Germany}
\address[HIRO]{Heidelberg Institute for Radiation Oncology -- HIRO, Im Neuenheimer Feld 280, 69120 Heidelberg, Germany}
\address{Department of Physics and Astronomy, Ruprecht Karls University Heidelberg, Grabengasse 1, 69117 Heidelberg, Germany}

\ead{n.wahl@dkfz.de}

\author{Philipp Hennig}
\address{Probabilistics Numerics, Max Planck Institute for Intelligent Systems, 72076 Tübingen, Germany}
\address{Chair for the Methods of Machine Learning, Eberhard Karls University Tübingen, 72024 Tübingen, Germany}

\author[DKFZ,HIRO]{Hans-Peter Wieser}
\address{Medical Faculty, Ruprecht Karls University Heidelberg, Grabengasse 1, 69117 Heidelberg, Germany}
\address{Department for Medical Physics, Ludwig-Maximilians-Universität München (LMU Munich), 85748 Garching b. München, Germany}

\author[DKFZ,HIRO]{Mark Bangert}

\version{2}

\renewcommand{\replaced}[2]{\added{#1}}
\renewcommand{\deleted}[1]{}

\makeatletter
\definecolor{changecolor}{RGB}{154,0,0}
\@namedef{Changes@AuthorColor}{changecolor}
\colorlet{Changes@Color}{changecolor}
\makeatother

\begin{document}
\frontmatter
\maketitle

\begin{abstract}
	\noindent\textbf{Purpose:} Radiotherapy, especially with charged particles, is sensitive to executional and preparational uncertainties that propagate to uncertainty in dose and plan quality indicators, e.\,g., \acp{DVH}. Current approaches to quantify and mitigate such uncertainties rely on explicitly computed error scenarios and are thus subject to statistical uncertainty and limitations regarding the underlying uncertainty model. Here we present an alternative, analytical method to approximate moments, in particular expectation value and (co)variance, of the probability distribution of \ac{DVH}-points, and evaluate its accuracy on patient data.\\
	\textbf{Methods:} We use \ac{APM} to derive moments of the probability distribution over individual \ac{DVH}-points based on the probability distribution over dose. By using the computed moments to parameterize distinct probability distributions over \ac{DVH}-points (here normal or beta distributions), not only the moments but also percentiles, \ie $\alpha$-\acp{DVH}, are computed. The model is subsequently evaluated on three patient cases (intracranial, paraspinal, prostate) in 30- and single-fraction scenarios by assuming the dose to follow a multivariate normal distribution, whose moments are computed in closed-form with \ac{APM}. The results are compared to a benchmark based on discrete random sampling. 
	\\ 
	\textbf{Results:} The evaluation of the new probabilistic model on the three patient cases against a sampling benchmark proves its correctness under perfect assumptions as well as good agreement in realistic conditions. More precisely, ca.~\SI{90}{\percent} of all computed expected \ac{DVH}-points and their standard deviations agree within \SI{1}{\percent} volume with their empirical counterpart from sampling computations, for both fractionated and single fraction treatments. When computing $\alpha$-\acp{DVH}, the assumption of a beta distribution achieved better agreement with empirical percentiles than the assumption of a normal distribution: While in both cases probabilities locally showed large deviations (up to \num{+-0.2}), the respective $\alpha$-\acp{DVH} for $\alpha = \{0.05,0.5,0.95\}$ only showed small deviations in respective volume (up to \SI{+-5}{\percent} volume for a normal distribution, and up to \SI{2}{\percent} for a beta distribution). A previously published model \replaced{from literature}{by different authors}, which was included for comparison, \replaced{exhibited substantially larger deviations}{did not yield reasonable $\alpha$-\acp{DVH}}.		
	\\
	\textbf{Conclusions:} With \ac{APM} we could derive a mathematically exact description of moments of probability distributions over \ac{DVH}-points given a probability distribution over dose. The model generalizes previous attempts and performs well for both choices of probability distributions, i.\,e., normal or beta distributions, over \ac{DVH}-points.  
\end{abstract}

\vspace{2pc}
\cleardoublepage

\mainmatter

\section{Introduction}
\label{sec:introduction}
Recent years have shown an increased interest in adequate, \replaced{patient}{case}-specific \deleted{plan} uncertainty quantification and mitigation for radiotherapy treatment planning both academically and clinically \parencite[see, e.\,g.,][and references therein]{Unkelbach2018,Fredriksson2012}. This development is, on the one hand, driven by emerging irradiation with particles and their characteristic sensitivity to uncertainties \autocite{Lomax2008,Lomax2008b}. On the other hand, it is facilitated by fast-growing computational capabilities that enable the computation of multiple dose scenarios with acceptable overhead.

Error dose scenarios are either computed as
\begin{enumerate*}[label=(\arabic*)]
	\item worst case estimates, \ie extreme realizations of the input uncertainty model which are used for robust optimization \parencite[as performed and evaluated within][]{Pflugfelder2008,Lowe2016,Lowe2017,Fredriksson2011,Fredriksson2016,Casiraghi2013a,Liu2012}, or 
	\item random samples from the probability distribution parameterizing the input uncertainty model \autocite{Unkelbach2007,Unkelbach2009,Gordon2009,Gordon2010,Bohoslavsky2013,Mescher2017} for stochastic approaches. 
\end{enumerate*}
An explicit derivation of probabilistic models remains the exception \autocite{Bangert2013, Sobotta2010, Sobotta2012, Perko2016, Wahl2018}.

Consequently, the analysis of plan uncertainty is based on the derived worst-case dose distributions or \enquote{error bar}-distributions with their respective  histograms \autocite{Liu2012,McGowan2015,Lowe2016,Lowe2017}, or statistical moments \autocite{Park2013,Unkelbach2009,Perko2016,Bangert2013,Wahl2018} as well as percentiles \autocite{Gordon2009,Gordon2010,Mescher2017}, according to the very optimization method used in the overall  planning workflow. This use of empirical uncertainty estimates, however, exhibits limitations, in particular concerning statistical accuracy and the required recomputations during optimization due to the changing pencil-beam weights. Further, they conceal the inherent mathematical transformation from the input probability space (e.\,g. set-up and range uncertainties) to the probability distribution over dose and the respective \ac{QI}. This aggravates their use in retrospective analyses and puts restrictions on the choice of optimization method and objectives, because the sampling pipeline cannot be inverted and/or efficiently differentiated.

\replaced{Approaches which rigorously propagate the uncertainty may overcome these limitations. Since they are able to provide analytical objectives, constraints, and derivatives, they can introduce new mathematical simplifications and improve computational efficiency by exploiting the analytical formulation for both uncertainty quantification and \added{mitigation with} probabilistic optimization approaches \autocite{Bangert2013,Wahl2018}. Because such approaches are not susceptible to statistical errors and the \enquote{curse of dimensionality} in scenario-based approaches, they deliver stable and reproducible results also for complex uncertainty models. However, derivation of such models is not trivial:}{While approaches which explicitly model the uncertainty propagation mathematically may overcome these limitations, derivation of such models is not \mbox{trivial \autocite{Bangert2013,Wahl2018}}.} Even if a model for dose probability is available, it still needs to be propagated to the derived plan indicators by hand. Other approaches overcome this step by re-sampling based on the derived dose uncertainty model \autocite{Perko2016,Sobotta2012}. 

For \ac{DVH}-points, analytical computation of moments of the probability distribution, given a probability distribution over the dose, has been attempted before \autocite{Henriquez2008,Henriquez2008a,Henriquez2010}. However, \textcite{Henriquez2008,Henriquez2008a} only provide a model for the expected value of \ac{DVH}-points with an upper bound on the \acp{DVH}' standard deviation. Further, only simplified uncertainty models for the underlying dose distribution were assumed: while different shapes of the distributions were evaluated, correlations between voxels were not modeled, even though correlations having crucial impact on the higher moments of the depending probability distribution. 

To derive a full model including correlations, this work will consequently not build on previous attempts, but provide a fresh start to a general methodology to compute the $\nu$-th moments of the probability distribution over \ac{DVH}-points. The goal is to derive a generally applicable model for \ac{DVH}-probabilities allowing arbitrary assumptions on the probability distribution over the dose distribution.

To do so, first a closed-form description for the moments of the probability distribution of \ac{DVH}-points is derived. Then, these moments parameterize a probability distribution over the respective \ac{DVH}-point. To evaluate our approach, three patient cases are investigated with statistical reference computations (using a large number of random dose samples from the probability distribution over set-up and range errors in fractionated and non-fractionated treatments). Along the lines of our validation campaign, we illustrate the shortcomings of previous work and highlight where wide approximations regarding the correlation models as exercised by \textcite{Henriquez2010} render meaningful quantification of \ac{DVH}-probabilities impossible.




\section{Materials \& Methods}
\subsection{\texorpdfstring{\acp{DVH}}{\acsp{DVH}} under uncertainty}
\subsubsection{Nominal computation}
\acp{DVH} are cumulative histograms over the spatial dose distribution in a \ac{VOI} $v$, here expressed as vector $\vecbf{d} \in \mathbb{R}_+^{\numvox}$ with number of voxels $\numvox$. Hence, for any given dose parameter $\hat{d}$, a \ac{DVH}-point $\mathrm{DVH}(\hat{d};\vecbf{d}) \in [0,1]$ equals the fraction of the volume that receives \emph{at least} dose $\hat{d}$. It can be expressed as averaged Heaviside steps
\begin{equation}
\mathrm{DVH}(\hat{d};\vecbf{d}) = \frac{1}{\numvox}\sum_{i\in v} \Theta(d_i - \hat{d})\label{eq:nomDVH}\,,
\end{equation}
meaning that only voxels $i$ with $d_i \geq \hat{d}$ contribute to the sum which is normalized by the total voxel count $\numvox$ in $v$ and thus yielding a fractional volume. Note that (without loss of generality) we assumed that all voxels have similar volume.

\subsubsection{Uncertainty analysis of \texorpdfstring{\acp{DVH}}{\acsp{DVH}}}
Uncertainty analysis of \acp{DVH} is mostly performed on an empirical basis through computation of error dose scenarios \parencite[among others][]{Park2013,Wahl2018,Kraan2013,Perko2016,Gordon2009,Gordon2010,Moore2009,Mescher2017,Unkelbach2009,Lowe2016,Lomax2008,Fredriksson2012}. This enables the computation of a \ac{DVH} for each dose scenario (which can be either a worst-case scenario or a random sample), from which then worst-case estimates, empirical statistical moments as well as quantiles of the probability distribution over \ac{DVH}-points are derived.

For the purpose of this work, three forms of \enquote{statistical} \acp{DVH} will be of importance. First, uncertainty of a \ac{DVH} can be evaluated through the statistical moments of each \ac{DVH}-point, for example the expected/mean \ac{DVH} and its standard deviation, which can be empirically determined from $n_s$ random dose samples $\vecbf{d}_s$ as
\begin{align}
	\overline{\mathrm{DVH}}(\hat{d}) &= \frac{1}{n_s} \sum_{s=1}^{n_s} \mathrm{DVH}(\hat{d}; \vecbf{d}_s)\\
	\sigma_{\mathrm{DVH}(\hat{d})} &= \sqrt{\frac{1}{n_s - 1} \sum_{s=1}^{n_s}\left[\mathrm{DVH}(\hat{d}; \vecbf{d}_s) - \overline{\mathrm{DVH}}(\hat{d})\right]^2}\,.
\end{align}
Secondly we discuss \enquote{$\alpha$-\acp{DVH}}, which can be compactly expressed as 
\begin{equation}
\alpha\text{-\ac{DVH}}(\hat{d};\vecbf{d},\alpha) = v_\alpha \Leftrightarrow P(\mathrm{DVH}(\hat{d}; \vecbf{d}) \leq v_\alpha) = \alpha
\label{eq:pdvh}
\end{equation}
where $v_\alpha$ is the volume covered with a probability $P(\mathrm{DVH}(\hat{d}) \geq v_\alpha) = \alpha$. Thus $\alpha$-\acp{DVH} can be used to give percentiles of the probability distribution of each \ac{DVH}-point and, together with the corresponding (1-$\alpha$)-\ac{DVH}, the respective confidence intervals. $\alpha$-\acp{DVH} may be computed with the empirical marginal quantile functions for the respective \ac{DVH}-points. Alternatively, $\alpha$-\acp{DVH} are equal to iso-probability curves on the respective \ac{DVCM} as proposed by \textcite{Gordon2009}. Such a \ac{DVCM} assigns a probability of coverage of each possible volume fraction for any dose threshold $\hat{d}$ and can be defined as
\begin{equation}
	\mathrm{DVCM}(\hat{d},v;\vecbf{d}) = P(\mathrm{DVH}(\hat{d}; \vecbf{d}) \leq v) = F_{\mathrm{DVH}(\hat{d},\vecbf{d})}(v)\,,\label{eq:DVCM}
\end{equation}
where $F_{\mathrm{DVH}(\hat{d},\vecbf{d})}(v)$ is the \ac{CDF} of the probability distribution over the respective \ac{DVH}-point.
\Cref{eq:DVCM} can then be directly inserted into \cref{eq:pdvh} such that the respective $\alpha$-\ac{DVH} is now the iso-curve at $\mathrm{DVCM} = \alpha$.

Note that such  $\alpha$-\acp{DVH} or \acp{DVCM} do not yield confidences or probabilities for the \emph{full} \ac{DVH} but only over single \ac{DVH}-points (\ie they represent marginal quantiles and \acp{CDF}), and hence do not generally represent naturally occuring \ac{DVH}-scenarios. 

\subsection{Moments of the probability distribution over dose-volume histograms}
\subsubsection{Analytical integration}
\label{sec:integrationDVH}
If the probability distribution over the dose $\vecbf{d}$ has the multivariate \ac{CDF} $F_{\vecbf{d}}$, the $\nu$-th moment of the probability distribution of a transformation $I(\vecbf{d})$ can be computed via integration
\begin{subequations}
\label[equation]{eq:moment_integral}%
\begin{align}
	\Exp{I(\vecbf{d})^\nu} &= \int_{\mathbb{R}^\numvox}I(\vecbf{\tilde{d}})^\nu\de F_{\vecbf{d}}(\vecbf{\tilde{d}})\label{eq:moment_integral:cdf}\\
	&= \int_{\mathbb{R}^\numvox}I(\vecbf{\tilde{d}})^\nu f_{\vecbf{d}}(\vecbf{\tilde{d}}) \de \vecbf{\tilde{d}}\,.\label{eq:moment_integral:pdf}
\end{align}
\end{subequations}

Moments of the probability distribution over a \ac{DVH} may thus be explicitly calculated by solving \cref{eq:moment_integral} for $I(\vecbf{d}) = \mathrm{DVH}(\hat{d};\vecbf{d})$. For the first \added{non-central} moment, this yields
\begin{subequations}
\label[equation]{eq:exp_dvh}	
\begin{align}
	\Exp{\mathrm{DVH}(\hat{d};\vecbf{d})} &= \int_{\mathbb{R}^{\numvox}} \frac{1}{\numvox}\sum_{i \in v} \Theta(\tilde{d}_i - \hat{d}) f_{\vecbf{d}}(\vecbf{\tilde{d}}) \de \vecbf{\tilde{d}}\label{eq:exp_dvh:integral}	\\
											&= \frac{1}{\numvox}\sum_i \int_{-\infty}^{\infty} \Theta(\tilde{d}_i - \hat{d}) f_{d_i}(\tilde{d}_i)  \de \tilde{d}_i\label{eq:exp_dvh:pulled_sum}\\
											&= \frac{1}{\numvox}\sum_i \int_{\hat{d}}^{\infty} f_{d_i}(\tilde{d}_i) \de \tilde{d}_i\label{eq:exp_dvh:removed_theta}\\
											&= \frac{1}{\numvox}\sum_i \left[1-F_{d_i}(\hat{d})\right]\,.\label{eq:exp_dvh:result}
\end{align}
\end{subequations}

Similar steps lead to the mixed non-central moment $\Exp{\mathrm{DVH}(\hat{d}_p;\vecbf{d})\mathrm{DVH}(\hat{d}_q;\vecbf{d})}$:
\begin{subequations}
\begin{align}
	\Exp{\mathrm{DVH}(\hat{d}_p;\vecbf{d})\mathrm{DVH}(\hat{d}_q;\vecbf{d})} 
		&= \int_{\mathbb{R}^{\numvox}} \frac{1}{\numvox^2}\sum_{il \in v} \Theta(\tilde{d}_i - \hat{d}_p)\Theta(\tilde{d}_l - \hat{d}_q) f_{\vecbf{d}}(\hat{d}) \de \vecbf{\tilde{d}}\\
		&= \frac{1}{\numvox^2}\sum_{il \in v} \int_{\mathbb{R}^{2}}  \Theta(\tilde{d}_i - \hat{d}_p)\Theta(\tilde{d}_l - \hat{d}_q) f_{\vecbf{d}_{i;l}}(\vecbf{\tilde{d}}_{i;l}) \de \tilde{\vecbf{d}_{i;l}}\\
		&= \frac{1}{\numvox^2}\sum_{il \in v} \int\limits_{\hat{d}_p}^\infty \int\limits_{\hat{d}_q}^\infty f_{\vecbf{d}_{i;l}}(\vecbf{\tilde{d}}_{i;l}) \de \tilde{d}_l\de \tilde{d}_i\,. \label{eq:mixed_dvh:result}
\end{align}
\end{subequations}

For the second non-central moment with $p=q$, \ie $	\Exp{\mathrm{DVH}(\hat{d}_p;\vecbf{d})^2}$, \cref{eq:mixed_dvh:result} can be expressed with the marginal bivariate cumulative distribution function $F_{\vecbf{d}_{i;l}}$ as
\begin{equation}
	\Exp{\mathrm{DVH}(\hat{d};\vecbf{d})^2} = \frac{1}{\numvox^2}\sum_{il \in v} \left[1 - F_{\vecbf{d}_{i;l}}\left(\hat{d}\vecbf{1}_2\right)\right]\label{eq:2nd_dvh}\,,
\end{equation}
where $\vecbf{1}_2 = (1,1)^\transposed$.

Together \cref{eq:exp_dvh,eq:mixed_dvh:result,eq:2nd_dvh} then give the (co)variance of \ac{DVH}-points at $\hat{d}_p$ and $\hat{d}_q$ using $\Cov{x}{y} = \Exp{xy} - \Exp{x}\Exp{y}$, \ie
\begin{multline}
	\Cov{\mathrm{DVH}(\hat{d}_p;\vecbf{d})}{\mathrm{DVH}(\hat{d}_q;\vecbf{d})} \\
	= \Exp{\mathrm{DVH}(\hat{d}_p;\vecbf{d})\mathrm{DVH}(\hat{d}_q;\vecbf{d})} - \Exp{\mathrm{DVH}(\hat{d}_p;\vecbf{d})}\Exp{\mathrm{DVH}(\hat{d}_q;\vecbf{d})}
	\label{eq:cov_dvh}
\end{multline}
which, in case of the variance of a \ac{DVH}-point at $\hat{d}$, consequently reduces to
\begin{equation}
	\Var{\mathrm{DVH}(\hat{d};\vecbf{d})} = \Exp{\mathrm{DVH}(\hat{d};\vecbf{d})^2} - \Exp{\mathrm{DVH}(\hat{d};\vecbf{d})}^2\,.
	\label{eq:var_dvh}
\end{equation}

Hence, for each point of a \ac{DVH} for a \ac{VOI} $v$, \crefrange{eq:exp_dvh}{eq:var_dvh} allow explicit computation of the expected value and variance of a \ac{DVH} as well as the covariance between all \ac{DVH} points, valid for any probability distribution over the dose $\vecbf{d}$ in $v$ as long as its univariate and bivariate marginal \ac{CDF} can be evaluated. 

While not explicitly evaluated in this work, similar steps can be taken to compute higher moments to more accurately parameterize the underlying probability distribution. This requires an expansion of the respective power of the sums of Heaviside steps with the multinomial theorem and evaluation of multivariate probabilities of higher dimensionality. We provide such a generalization in \cref{app:multi}.

\subsubsection{Summary of previous work}
\label{sec:previous_works_expDVH}
\added{A result similar to \cref{eq:exp_dvh} was already documented in literature\autocite{Henriquez2008}, where it was obtained}
\deleted{{\Textcite{Henriquez2008}} proposed to analytically compute expected \ac{DVH} points} 
by interpreting the computation of the \ac{DVH} according to \cref{eq:nomDVH} as a sum of Bernoulli experiments: Each voxel $i$ falls into the current \added{\ac{DVH}} bin at $\hat{d}$ with a probability $p_i = P(d_i > \hat{d}) = 1-F_{d_i}(\hat{d})$, where $F_{d_i}$ is the marginal \ac{CDF} for $d_i$, and does not fall into the bin with a probability $1-p_i$. \added{With the linearity of the expectation value one then can directly derive \cref{eq:exp_dvh:result}.} 

\deleted{Because of the linearity of the expectation value, the expected \ac{DVH} point at $\hat{d}$ is then given as
\begin{equation}
\Exp{\mathrm{DVH}(\hat{d};\vecbf{d})} = \frac{1}{\numvox}\sum_{i \in v}\left[ 1-F_{d_i}(\hat{d})\right]\,,
\label{eq:exp_dvh_bernoulli}
\end{equation}
which corresponds to the result in \cref{eq:exp_dvh}.}

\added{The result proved to be applicable with different families of assumed probability distributions (\ie Gaussian, triangular and rectangular/uniform)\autocite{Henriquez2008,Henriquez2008a}. However, due to the simplified uncertainty model using constant relative standard deviation and no correlation between voxels, an exact computation of higher moments of a \ac{DVH}-point's probability distribution was not attempted.}

\deleted{{\Textcite{Henriquez2008,Henriquez2008a}} evaluated \cref{eq:exp_dvh_bernoulli} for different families (\ie Gaussian, triangular and rectangular/uniform) of probability distributions  over the respective voxel dose values. Since they did not rely on explicitly propagated uncertainties but only on nominal dose distributions, they set $\Exp{\vecbf{d}} \overset{!}{=} \vecbf{d}$ and $\sigma_{\vecbf{d}} \overset{!}{=} c\cdot \vecbf{d}$, \ie constant relative standard deviation. Due to this simplified uncertainty model lacking correlation, \textcite{Henriquez2008} did not attempt to compute higher moments like the variance.}

\subsection{Confidence bounds for \texorpdfstring{\ac{DVH}}{\acs{DVH}}-points}
\subsubsection{Parameterization of the \texorpdfstring{\ac{DVH}}{\acs{DVH}} probability distribution}
Since \crefrange{eq:exp_dvh}{eq:var_dvh} provide expected value and covariance of any \ac{DVH}-point, one could possibly directly parameterize the probability distribution over the full \ac{DVH} with a multivariate normal distribution. This parameterization is, however, unphysical; since \ac{DVH}-points represent the fraction of a volume, their values are confined to the interval $[0,1]$ in contrast to the infinite support of the multivariate normal distribution. Hence, the probability distribution of a \ac{DVH} point might be more \enquote{physically} represented by a distribution supported only in the interval $[0,1]$, such as a beta distribution $\mathcal{B}(a,b)$ with shape parameters $a$ and $b$. The beta distribution is shortly characterized in \cref{app:beta}. However, lacking a generalized multivariate form \parencite[for recent approaches on constructing bivariate beta distributions see, e.\,g.,][]{Olkin2015,Nadarajah2017}, $\mathcal{B}(a,b)$ may only be used to parameterize the marginal distribution over a single \ac{DVH}-point, not the full multivariate \ac{DVH}. 

Current approaches \replaced{applying}{that working with} \ac{DVH} confidences define these on a marginal by-point basis \autocite{Mescher2017,Gordon2009,Gordon2010,Moore2009}. Hence, quantifying probabilities over marginal \ac{DVH}-points is in line with literature and enables comparability. Therefore, in this work, marginal probabilities will be evaluated, based on the (unphysical) parameterization with normal distributions as well as the more physical approach using a beta distribution whose shape parameters $a$ and $b$ are obtained from the respective \ac{DVH}-point's expectation value and variance with \cref{eq:app:beta:alpha_est,eq:app:beta:beta_est}.

Using either a normal or beta distribution, one can directly compute \acp{DVCM} according to \cref{eq:DVCM} or an $\alpha$-\ac{DVH} using 
\begin{equation}
\begin{aligned}
	\alpha\text{-DVH}(\hat{d};\vecbf{d}, \alpha) &= F^{-1}_{\mathrm{DVH}(\hat{d};\vecbf{d})}(\alpha) \\
	&= \begin{cases} \Exp{\mathrm{DVH}(\hat{d};\vecbf{d})} + \sqrt{2\Var{\mathrm{DVH}(\hat{d};\vecbf{d})}} \operatorname{erf}^{-1}(2\alpha - 1) & \text{| normal} \\
	I^{-1}_\alpha(a,b) & \text{| beta} \end{cases}\,,
\end{aligned}
\label{eq:pdvh_quantile}
\end{equation}
where $\operatorname{erf}^{-1}$ denotes the inverse error function and $I^{-1}_\alpha$ represents the inverse of the regularized incomplete beta-function.

\subsubsection{Summary of previous work}
\label{sec:previous_works_pdvh}
\added{A subsequent work\autocite{Henriquez2010} to \textcite*{Henriquez2008,Henriquez2008a}}
\deleted{In a subsequent work to \textcite*{Henriquez2008,Henriquez2008a}, \textcite{Henriquez2010} attempt to derive confidence intervals for \ac{DVH}-points based on the}
\added{already formulated an analyitcal}
calculation of $\alpha$-\acp{DVH} as defined in \cref{eq:pdvh}. 
\deleted{However, they define $\alpha$-\ac{DVH}-points as the volume \textcquote{Henriquez2010}{receiving a dose equal to or greater than \textins{$\hat{d}$} with a certainty equal or greater than 1-$\alpha$}. This definition is not fully correct, since the respective certainty must be only equal to (and not greater than) $1-\alpha$ to be consistent with their following derivations using \acp{CDF}: First, they define a}
In summary, their model is based on the definition of a  \textcquote{Henriquez2010}{binary random variable}
\begin{equation}
T_i^{\alpha,\hat{d}} = \begin{cases} 1 & P(d_i \geq \hat{d}) > 1 - \alpha\\ 0 & P(d_i \geq \hat{d}) \leq 1 - \alpha \end{cases} \label{eq:binary_hc}
\end{equation}
interpreted as \textcquote{Henriquez2010}{the volume receiving a dose greater than \textins{$\hat{d}$} with a probability greater than $1-\alpha$}. This interpretation then leads them to define $T_i^{\alpha,\hat{d}} = \Theta\left(1 - \alpha - F_{d_i}(\hat{d})\right)$ which, in analogy to \cref{eq:pdvh}, translates to
\begin{equation}
\alpha\text{-DVH}^{\text{HC}}(\hat{d};\vecbf{d}, \alpha) = \frac{1}{\numvox}\sum_{i \in v} T_i^{\alpha,\hat{d}} = \frac{1}{\numvox}\sum_{i \in v} \Theta\left(1 - \alpha - F_{d_i}(\hat{d})\right)\,.\label{eq:pdvh_hc}
\end{equation}

\Cref{eq:pdvh_hc} substantially differs from our result in \cref{eq:pdvh_quantile}\deleted{using the quantile function of a probability distribution  parameterized with their moments, which in return are obtained by evaluating \cref{eq:var_dvh,eq:exp_dvh}}. \added{This may be attributed to disregarding correlation across voxels in their derivations or inconsistencies in the definition of $T_i^{\alpha,\hat{d}}$, which apparently does not} 
\deleted{\Cref{eq:pdvh_hc} does not factor in the various possible correlations between voxels. Further, the definition of $T_i^{\alpha,\hat{d}}$ is unclear, since it does not} describe a random variable per se, but rather includes the evaluation of a probability---in this case the probability $P(d_i - \hat{d})$ of uncertain dose $d_i$ exceeding $\hat{d}$---which is not a random but a fixed value obtained from the \ac{CDF} over $d_i$. \deleted{These points of criticism question the validity of \cref{eq:pdvh_hc} for quantification of reasonable confidences over \ac{DVH}-points under uncertainty. Since \textcite{Henriquez2010} did also not validate \cref{eq:pdvh_hc} against statistical estimates from sampling, we include \cref{eq:pdvh_hc} in our analysis to further explore the implications of the simplifications in the derivation of said \namecref{eq:pdvh_hc}.}
\added{Since \cref{eq:pdvh_hc} has only been tested with an assumed uncertainty model and not been benchmarked against sample statistics\autocite{Henriquez2010}, we incorporate \cref{eq:pdvh_hc} in our evaluations and discuss the \namecref{eq:pdvh_hc} further in \cref{sec:discussion}.}

\subsection{Dose Uncertainty Model}
\label{sec:doseunc}
Evaluation of \cref{eq:exp_dvh,eq:var_dvh,eq:cov_dvh} or, in general, \cref{eq:app:multiindex_Dvh} requires a model for the probability distribution over dose $\vecbf{d}$, such that its \ac{CDF} can be evaluated. Note that empirical \acp{CDF} as well as analytical/parameterized \acp{CDF} can be used.  
\subsubsection{Gaussian model for the probability distribution over dose}
\label{sec:gaussian_model}
For a first evaluation and validation of the new probabilistic computations, we assume
\begin{equation}
\vecbf{d} \sim \mathcal{N}(\vecbf{\mu},\Sigma)\,,
\label{eq:dose_uc_model}
\end{equation}
\ie the dose follows a multivariate normal distribution with mean dose $\vecbf{\mu}$ and covariance $\Sigma$. This choice of probability distribution cannot represent the \emph{true} underlying probability distribution: First, the multivariate normal is supported on the full multidimensional real space, while physical dose is bound to the positive orthant. And second, empirical evidence \parencite[\eg][]{Park2013,Lowe2016} as well as heuristic considerations show that the respective distribution exhibits considerable skewness and is consequently not part of the symmetric Gaussian family. Alternatives to assumption \eqref{eq:dose_uc_model} will be discussed in \cref{sec:discussion}. As a first order approximation, however, \cref{eq:dose_uc_model} is well suited to study the probabilistic \ac{DVH}-model, because on the one hand, its univariate and bivariate probabilities can be calculated \autocite{Genz2004}, which is sufficient to compute an expected \ac{DVH} and its (co)variance. On the other hand, evaluation on patient cases with assumption \eqref{eq:dose_uc_model} implicitly studies the impact of the inaccurate Gaussian dose model on the evaluation of \ac{VOI}-based dose statistics like \acp{DVH} under uncertainty. Further, while for a single fraction treatment the non-Gaussian shape of the probability distribution over dose is to be expected, under multiple fractions a more Gaussian-like shape may form \parencite[for examples of voxel dose probability distributions see][]{Park2013,Lowe2016}.

\subsubsection{Computation of dose uncertainty}
\label{sec:apm}
The Gaussian dose model from \cref{sec:gaussian_model} requires mean $\vecbf{\mu}$ and covariance $\Sigma$ for evaluation. E.\,g., these could be empirically estimated with sample mean and covariance of a set of discrete error scenarios.

Since the motivation of this work is to build a fully analytical model (which also facilitates future use in optimization), we rely on computation of $\vecbf{\mu}$ and $\Sigma$ through \ac{APM} as introduced by \textcite{Bangert2013}. In previous works we could already show that \ac{APM} accurately models $\vecbf{\mu}$ and $\vecbf{\sigma} = \sqrt{\operatorname{diag}(\Sigma)}$ \autocite{Wahl2017a}, and that efficient application to patient data---especially in the context of fractionation \autocite{Wahl2018}---is possible. \Textcite{Wieser2017b} further extended it to biological optimization by demonstrating \acp{APM} applicability to intensity-modulated carbon ion therapy planning.

\ac{APM} acts as a probabilistic pencil-beam dose calculation algorithm inherently enabling computation of moments of the probability distribution over the resulting dose. More exact, \ac{APM} represents the constituents of a pencil-beam algorithm as superpositions of Gaussian functions (including the integrated depth dose, \ie the Bragg peak), enabling propagation of uncertainties through the dose calculation in closed-form via analytical integration \parencite[for a detailed explanation see][]{Bangert2013}.

The accuracy of this approximation of the pencil-beam algorithm can, in principle, be arbitrarily chosen by varying the number of Gaussian components, thus providing also a nominal dose calculation algorithm for inverse treatment planning that is similar in quality to common pencil-beam algorithms. As such, it is able to provide a dose influence matrix $D \in \mathbb{R}_+^{\numvox \times \numbix}$ with number of voxels \numvox\xspace and number of pencil-beams \numbix\xspace  generating the dose $\vecbf{d} \in \mathbb{R}_+^{\numvox}$ from the fluence vector $\vecbf{w} \in \mathbb{R}_+^{\numbix}$ via the linear transformation 
\begin{equation}
d_i = \sum_{j}D_{ij}w_j\,.
\label{eq:dose_map}
\end{equation}
In \cref{eq:dose_map}, $i$ indexes voxels in the patient while $j$ indexes pencil-beams.

Now, in addition to \cref{eq:dose_map}, \ac{APM} provides probabilistic analogs to \cref{eq:dose_map} for moments of the probability distribution over $\vecbf{d}$ by enabling element-wise computation of expectation value and (co)variance of elements of the dose influence matrix $D$. This allows to represent the expected value of dose $\Exp{\vecbf{d}}$ as a linear transformation
\begin{equation}
	\Exp{d_i} = \sum_j\underbrace{\Exp{D_{ij}}}_{\ExpDij} w_j = \sum_j \ExpDij w_j
\end{equation}
and the covariance in dose with a quadratic form
\begin{equation}
\Cov{d_i}{d_l} = \sum_{jm}\underbrace{\Cov{D_{ij}}{D_{lm}}}_{\CovInfluence} w_j w_m = \sum_{jm} \CovInfluence  w_j w_m\,.
\end{equation}
Hence, one can denote $\ExpD \in \mathbb{R}_+^{\numvox \times \numbix}$ and $\Vtensor \in \mathbb{R}_+^{\numvox \times \numbix \times \numvox \times \numbix}$ as \emph{expected dose influence matrix} and \emph{covariance influence tensor}, respectively. While $\Vtensor$ is, in general, too large to be stored in memory, the element-wise computation with \ac{APM} allows on-the-fly evaluation of dependent quantities, \eg the variance or covariance of dose.

\subsection{Validation and application of the model}
The analytical probabilistic \ac{DVH}-model will be evaluated on three patient cases, \ie an intracranial, a paraspinal, and a prostate case. Parameters used for planning and uncertainty computations are laid out in \cref{tab:patient_data}.  These cases were already evaluated in our previous works \autocite{Wahl2017a,Wahl2018,Wieser2017b}. Furthermore, a detailed comparison of $\alpha$-\acp{DVH} and \acp{DVCM} is exercised to validate empirical percentiles against results from the respective quantile functions from \cref{eq:pdvh} and the previous works by \textcite{Henriquez2010}. 

\subsubsection{Application on all cases using the fractionated treatment samples}

For all three patient cases, empirical estimates $\overline{\mathrm{DVH}}$, $\bar{\sigma}^2_{\mathrm{DVH}}$ are obtained by computing $\num{100}\times\num{30}$ dose scenarios, \ie \num{100} sampled treatment scenarios with \num{30} fraction each, based on a Gaussian uncertainty model using the assumed setup and range errors from \cref{tab:patient_data} \added{as respective standard deviation with zero mean \parencite[compare to the description of the \ac{APM} uncertainty model in][]{Bangert2013}}. Additionally, expected dose $\vecbf{\mu}$ and covariance $\Sigma$ within the respective \acp{VOI} was also computed with \ac{APM} (see \cref{sec:apm}) for the same input uncertainty model, and then fed into the herein presented \ac{DVH}-models (by assuming a multivariate normal distribution as described in \cref{sec:doseunc}). This enabled comparison of the sample statistics to a fully analytical method and serves as proof-of-concept of the derived model.

\subsubsection{Full validation on the intracranial case with \num{5000} samples}
As the intracranial case is the smallest one with lowest computational overhead, we further computed \num{5000} realizations of single-fraction treatments. These will be used to nearly eliminate statistical inaccuracy for benchmarking \ac{APM}. 

To further validate the analytical computation itself excluding inherent mismatch of modeled and real probability distribution over dose, the analytically computed $\vecbf{\mu}$ and $\Sigma$ are additionally used to create \num{5000} new dose samples {\emph{under the assumption that dose \emph{actually} follows a multivariate normal distribution}, \ie their samples are drawn from the distribution $\mathcal{N}(\vecbf{\mu},\Sigma)$. From these samples, a second statistical estimate of the \ac{DVH} is obtained. This can be used to validate if the analytical computation is actually correct under the multivariate normal assumption from \cref{eq:dose_uc_model}. Further, \acp{DVCM} and $\alpha$-\acp{DVH} are computed under the assumption of marginally normally distributed \ac{DVH}-points and marginally beta distributed \ac{DVH}-points. These are compared to the respective statistical estimates from the \num{5000} scenario samples.


\section{Results}
We evaluated the described methodology on three patient cases -- an intracranial, paraspinal, and prostate patient. Information about the datasets, treatment plans and the assumed input uncertainty model can be found in \cref{tab:patient_data}.

\subsection{Proof of work -- computation on patient data}
\Cref{fig:probdvhs} compares sample mean and standard deviation of \acp{DVH} to the respective analytical computations with \crefrange{eq:exp_dvh}{eq:var_dvh} for treatments with \num{30} fractions.

\begin{figure}[p]
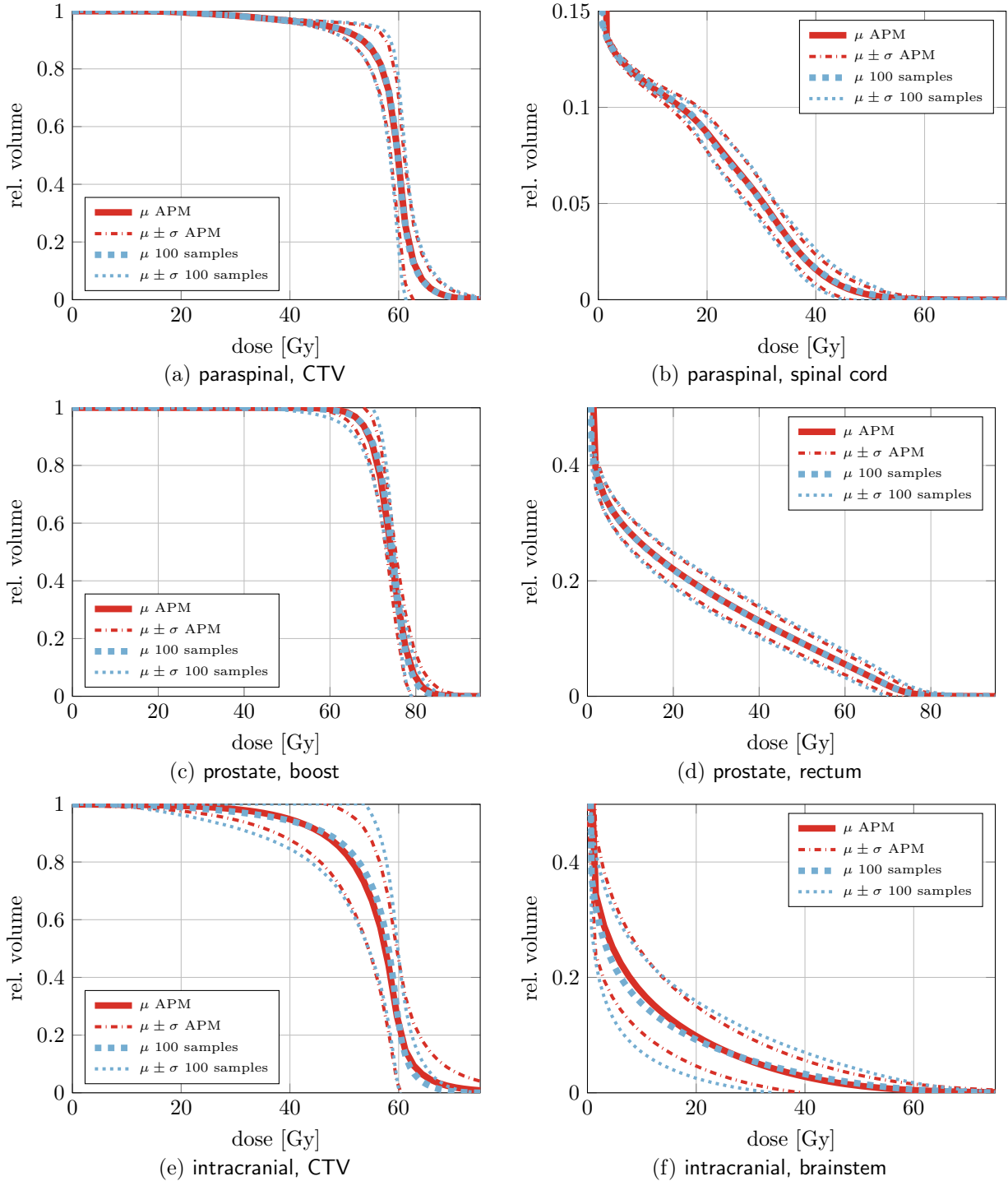

	\footnotesize
	\setlength\subfigwidth{0.5\textwidth}
	\setlength\figurewidth{0.85\subfigwidth}
	\setlength\figureheight{0.7\figurewidth}
	\begin{subfigure}[b]{\subfigwidth}
		\includetikz{probDVH_spinal_CTV_30frac}
		\caption{paraspinal, \ac{CTV}}
	\end{subfigure}
	\begin{subfigure}[b]{\subfigwidth}
		\includetikz{probDVH_spinal_spinalcord_30frac}
		\caption{paraspinal, spinal cord}
	\end{subfigure}
	
	\begin{subfigure}[b]{\subfigwidth}
		\includetikz{probDVH_prostate_boost_30frac}
		\caption{prostate, boost}
	\end{subfigure}
	\begin{subfigure}[b]{\subfigwidth}
		\includetikz{probDVH_prostate_rectum_30frac}
		\caption{prostate, rectum}
	\end{subfigure}

		\begin{subfigure}[b]{\subfigwidth}
		\includetikz{probDVH_head_CTV_30frac}
		\caption{intracranial, \ac{CTV}}
	\end{subfigure}
	\begin{subfigure}[b]{\subfigwidth}
		\includetikz{probDVH_head_brainstem_30frac}
		\caption{intracranial, brainstem}
	\end{subfigure}
	\caption{Analytically computed expectation value and standard deviation of \acp{DVH} of a target volume and \ac{OAR} for each of the three patient cases, compared to the respective sample mean and standard deviation. For the sampling benchmark, \num{100} treatments were simulated by multivariate normal sampling using the systematic errors from \cref{tab:patient_data} as standard deviation, while for each treatment taking \num{30} fraction samples based on the random component.}
	\label{fig:probdvhs}
\end{figure}

For the prostate case, the sampled and analytically computed mean \ac{DVH} and its standard deviation yield good agreement for both target and \ac{OAR}. For the intracranial case, especially the curves illustrating standard deviation seem to exhibit larger differences. However, a closer look reveals that the differences originate from both discrepancies of the mean and the standard deviation estimates.

To better quantify the differences between the analytical computation and the sample reference, \cref{fig:dev_histograms} summarizes the absolute difference in relative volume for all patients grouped by
\begin{enumerate*}[label=(\arabic*)]
	\item mean and standard deviation,
	\item targets and OARs, and
	\item 1 and 30 fraction treatments.	
\end{enumerate*}  Differences between analytical and sample computations are, in general, larger for targets than for \acp{OAR}. For the \acp{OAR}, the evaluation for multiple fractions shows an increase in accuracy. This does not seem to transfer to targets, where, although the number of points with minimal difference ($<\num{0.001}$) also increases, a stronger tail to higher differences is present. In general, more outliers, \ie single \ac{DVH}-points with large difference, can be observed when performing the calculations for a treatment in 30 fractions.

\begin{figure}[htb]
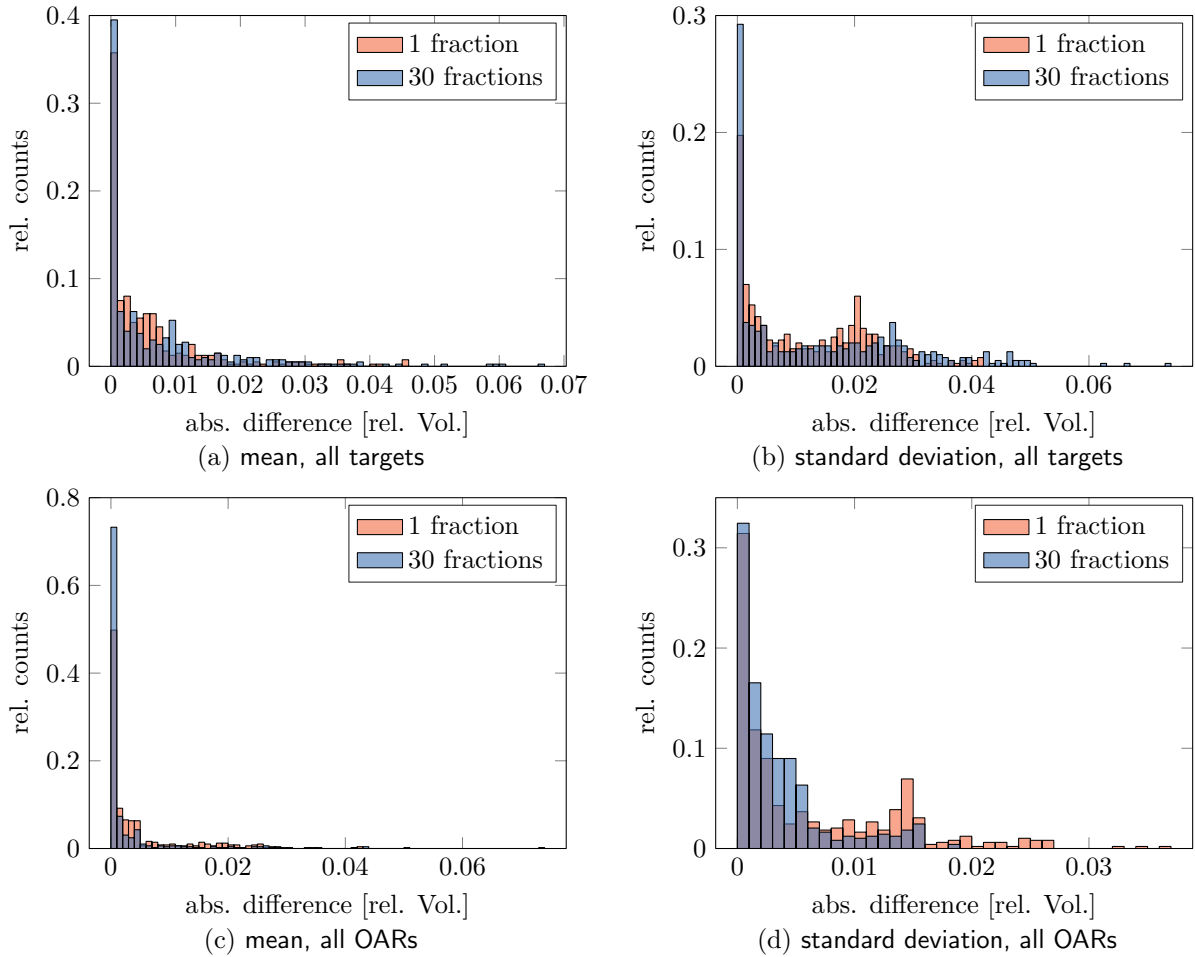

	\footnotesize
	\setlength\subfigwidth{0.5\textwidth}
	\setlength\figurewidth{0.85\subfigwidth}
	\setlength\figureheight{0.7\figurewidth}
	\begin{subfigure}[b]{\subfigwidth}
		\includetikz{absDiffHist_head_prostate_spinal_mean_targets}
		\caption{mean, all targets}
	\end{subfigure}
	\begin{subfigure}[b]{\subfigwidth}
		\includetikz{absDiffHist_head_prostate_spinal_std_targets}
		\caption{standard deviation, all targets}
	\end{subfigure}
	
	\begin{subfigure}[b]{\subfigwidth}
		\includetikz{absDiffHist_head_prostate_spinal_mean_oars}
		\caption{mean, all \acp{OAR}}
	\end{subfigure}
	\begin{subfigure}[b]{\subfigwidth}
		\includetikz{absDiffHist_head_prostate_spinal_std_oars}
		\caption{standard deviation, all \acp{OAR}}
	\end{subfigure}
	\caption{Histograms (bin width: \num{0.001}) of the absolute differences observed between the mean and standard deviation of all \ac{DVH}-points computed analytically and via random sampling for all patients. (a,b) show the analysis for all evaluated targets in all patients, while (c,d) display it for all evaluated \acp{OAR}.}
	\label{fig:dev_histograms}
\end{figure}

\subsection{Validation of model and analytical computations}
\subsubsection{Distribution of single \texorpdfstring{\ac{DVH}}{\acs{DVH}}-points}
\label{sec:dvh_points_validate}
\Cref{fig:dvh_points_validated} shows normalized histograms of two \ac{DVH}-points (one evaluated at \SI{57}{\gray} for the target and one at \SI{30}{\gray} for the brainstem of the intracranial case), comparing the samples from the dose scenarios and from the multivariate normal approximation. Their respective approximations with a normal distribution visualize differences between the respective moments of the \ac{DVH}-point's probability distribution: in the \ac{CTV}, the re-sampled mean underestimates the \ac{DVH} at \SI{57}{\gray} by a volume fraction of \SI{4}{\percent}, whereas in the brainstem difference between the mean / expected values is negligible. The opposite holds true for the computed standard deviation. 

\begin{figure}[htb]
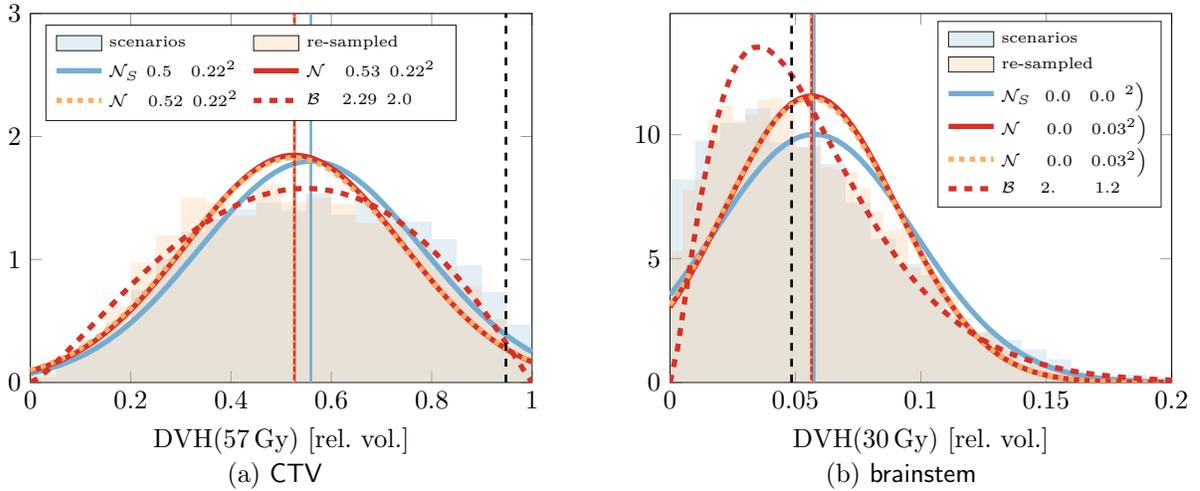

	\footnotesize
	\setlength\subfigwidth{0.5\textwidth}
	\setlength\figurewidth{0.85\subfigwidth}
	\setlength\figureheight{0.7\figurewidth}
	\begin{subfigure}[b]{\subfigwidth}
		\includetikz{qiValidation_dvP_CTV}
		\caption{\ac{CTV}}
		\label{fig:dvh_points_validated:ctv}
	\end{subfigure}
	\begin{subfigure}[b]{\subfigwidth}
		\includetikz{qiValidation_dvP_HIRNSTAMM}
		\caption{brainstem}
		\label{fig:dvh_points_validated:brainstem}
	\end{subfigure}
	\caption{Probability distribution over \ac{DVH}-points evaluated at $\hat{d} = \SI{57}{\gray}$ in the \ac{CTV} (a) and at \SI{30}{\gray} in the brainstem (b) of the intracranial case. The histograms show the distribution from the \num{5000} dose scenarios and the \num{5000} dose distributions  re-sampled under assumption \eqref{eq:dose_uc_model}. $\mathcal{N}_\mathcal{S}$ represents a normal distribution parameterized from sample mean and variance from the \num{5000} \acp{DVH} obtained from the scenario samples, $\mathcal{N}_\mathcal{R}$ has ben similarly computed from the re-sampled scenarios. $\mathcal{N}_\mathcal{A}$ parameterizes a normal distribution based on the analytical (\ac{APM}) computation of \ac{DVH}-point expectation and variance, and $\mathcal{B}_\mathcal{A}$ uses the same values to parameterize a beta distribution. The vertical lines indicate the respective expected/mean values, with the dashed black line giving the nominal value.}
	\label{fig:dvh_points_validated}
\end{figure}

The analytically computed expectation value and variance of the respective \ac{DVH}-points shows no significant difference to the statistical moments obtained from the re-sampled data. This is expected, since the analytical computations are mathematically exact and only negligible numerical inaccuracy is introduced when evaluating the univariate and bivariate normal \ac{CDF}. 

The Gaussian approximation is not bound to the volumetric interval $[0,1]$, and thus would assign non-zero probability to non-existing, e.\,g., negative, volume fractions. \Cref{fig:dvh_points_validated} thus shows corresponding approximations with beta distributions, whose shape parameters are obtained from \cref{eq:app:beta:alpha_est,eq:app:beta:beta_est} using analytically computed expectation and variance of the \ac{DVH}-point. This leads to a physically more reasonable distribution which is further backed by the Q-Q plots comparing the quantiles of the normal and beta approximation to the empirical quantiles in \cref{fig:qqplot}.

\begin{figure}[htb]
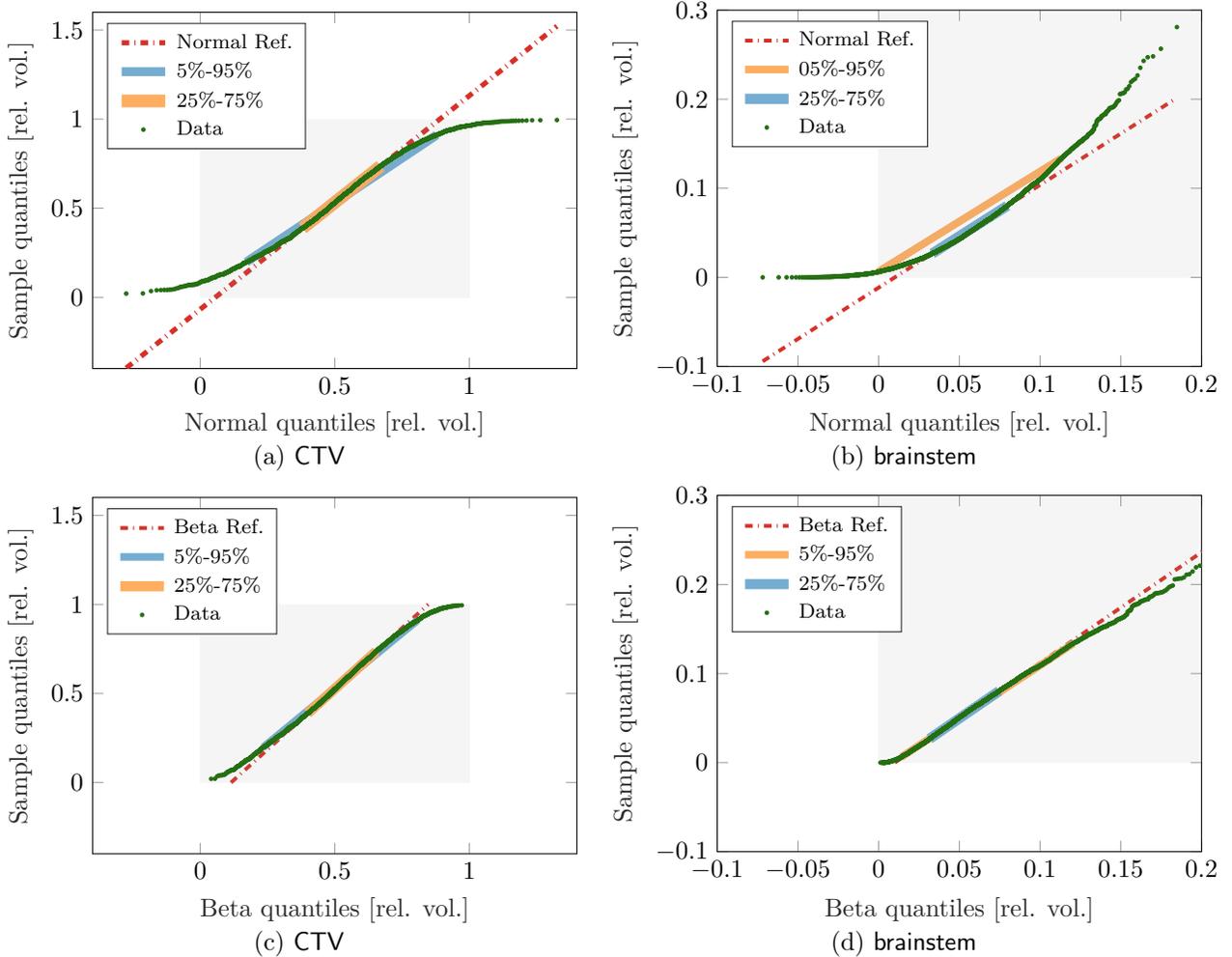

	\footnotesize
	\setlength\subfigwidth{0.5\textwidth}
	\setlength\figurewidth{0.85\subfigwidth}
	\setlength\figureheight{0.7\figurewidth}
	\begin{subfigure}[b]{\subfigwidth}
		\includetikz{qqPlot_dv_normal_CTV}
		\caption{\ac{CTV}}
		\label{fig:qqplot:normal_CTV}
	\end{subfigure}
	\begin{subfigure}[b]{\subfigwidth}
		\includetikz{qqPlot_dv_normal_HIRNSTAMM}
		\caption{brainstem}
		\label{fig:qqplot:normal_brainstem}
	\end{subfigure}

	\begin{subfigure}[b]{\subfigwidth}
		\includetikz{qqPlot_dv_beta_CTV}
		\caption{\ac{CTV}}
	\end{subfigure}
	\begin{subfigure}[b]{\subfigwidth}
		\includetikz{qqPlot_dv_beta_HIRNSTAMM}
		\caption{brainstem}
		\label{fig:qqplot:beta_brainstem}
	\end{subfigure}
	\caption{Quantile-quantile plots comparing empirical quantiles (y-axis) for the data (green) to quantiles from the hypothesized normal (a,b) and beta (c,d) distributions (x-axis) obtained from analytical moment computations (red). The thicker green and yellow lines span the \SIrange[range-phrase={--}]{5}{95}{\percent} and \SIrange[range-phrase={--}]{25}{75}{\percent} quantiles, respectively. In each plot the gray area additionally enclose \enquote{physically feasible} volumes in $[0,1]$. \ac{CTV} and brainstem of the intracranial case are shown for the same \ac{DVH}-points as in \cref{fig:dvh_points_validated}. The data is based on the \num{5000} dose scenario samples for a single fraction.}
	\label{fig:qqplot}
\end{figure}

\Cref{fig:qqplot:normal_CTV,fig:qqplot:normal_brainstem} underline the problem of the infinite support of the normal distribution, \ie unphysical quantiles exist in the theoretical normal model. This goes hand in hand with an especially pronounced disagreement between theoretical and empirical quantiles approaching the boundaries of the interval $[0,1]$, but also concerns possible skewness (compare \cref{fig:qqplot:normal_brainstem,fig:dvh_points_validated:brainstem}) or excess kurtosis of the distribution (\cref{fig:qqplot:normal_CTV,fig:dvh_points_validated:ctv}). These disagreements are reduced using a beta distribution. Especially the evaluated \ac{DVH}-point in the brainstem in \cref{fig:qqplot:beta_brainstem} shows near perfect agreement with the empirical quantiles. For all four evaluations in \cref{fig:qqplot}, however, near perfect agreement is achieved for \enquote{inner} quantiles, \ie the first and third quartile. 

\subsubsection{Evaluation of cumulative probabilities}
Next, we assess the accuracy of complete $\alpha$-\acp{DVH} by comparing $\alpha$-\acp{DVH} computed from the quantile functions of the respective probability distribution parameterized from the analytical computations with empirical quantiles over the full \ac{DVH}. Furthermore we compare to the previous attempt of analytical computation of $\alpha$-\acp{DVH} \autocite{Henriquez2010} as laid out in \cref{sec:previous_works_pdvh}.

\Cref{fig:pdvhs} shows the respective comparisons of analytically computed \acp{DVCM} to \acp{DVCM} obtained from sample statistics (compare \cref{eq:DVCM}).
The $\alpha$-\acp{DVH} in \cref{fig:pdvhs:ctv_ch,fig:pdvhs:brainstem_ch} computed with \cref{eq:pdvh_hc}, \ie the method from \textcite{Henriquez2010}, show significant differences to the corresponding reference $\alpha$-\acp{DVH} from sampling. 

\begin{figure}[p]
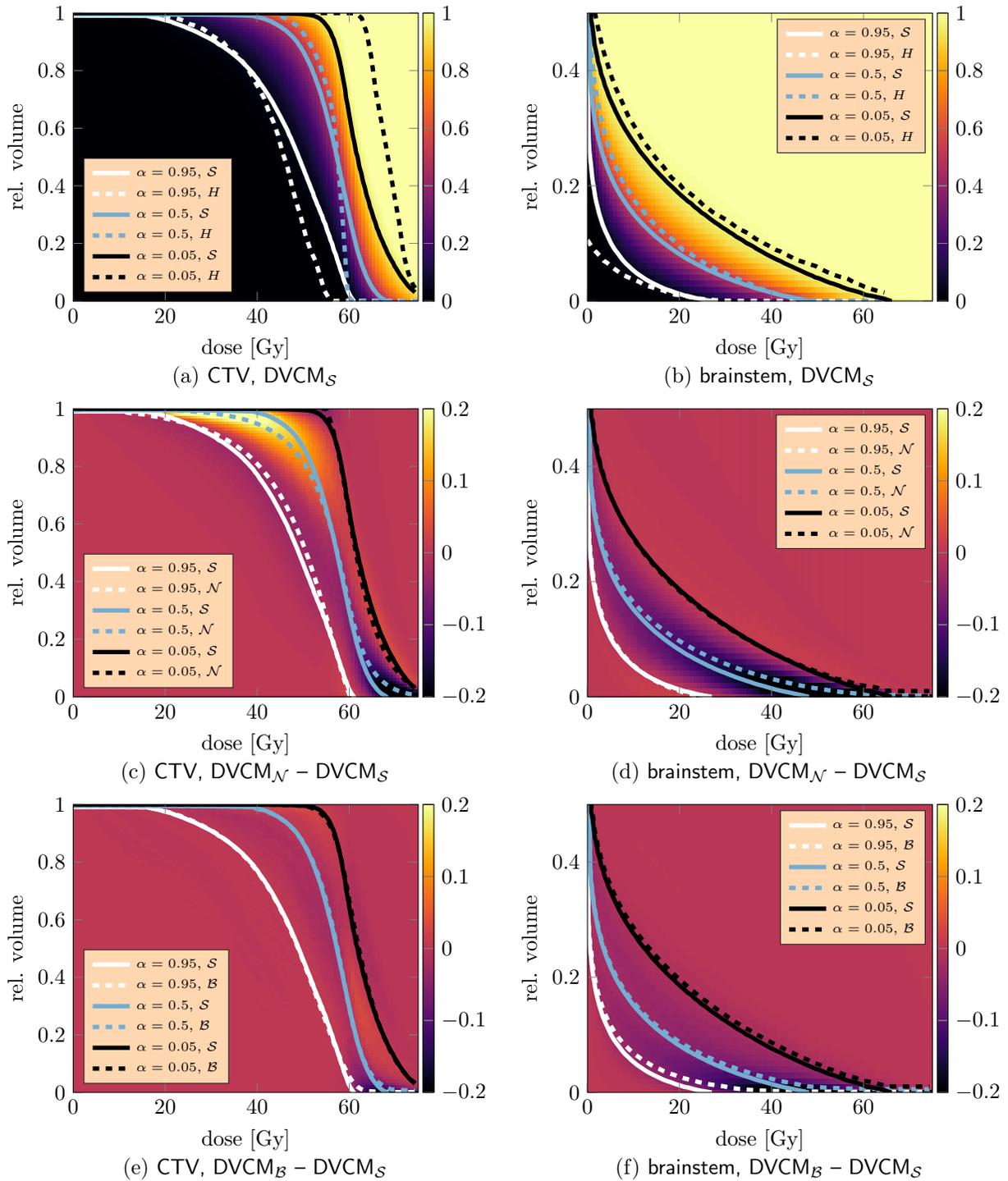

	\footnotesize
	\setlength\subfigwidth{0.5\textwidth}
	\setlength\figurewidth{0.85\subfigwidth}
	\setlength\figureheight{0.7\figurewidth}
	\begin{subfigure}[b]{\subfigwidth}
		\includetikz{dvcm_CTV}
		\caption{\ac{CTV}, \ac{DVCM}$_{\mathcal{S}}$}
		\label{fig:pdvhs:ctv_ch}
	\end{subfigure}
	\begin{subfigure}[b]{\subfigwidth}
		\includetikz{dvcm_brainstem}
		\caption{brainstem, \ac{DVCM}$_{\mathcal{S}}$}
		\label{fig:pdvhs:brainstem_ch}
	\end{subfigure}
	
	\begin{subfigure}[b]{\subfigwidth}
		\includetikz{dvcm_gaussDiff_CTV}
		\caption{\ac{CTV}, \ac{DVCM}$_{\mathcal{N}}$ -- \ac{DVCM}$_{\mathcal{S}}$}
		\label{fig:pdvhs:ctv_n}
	\end{subfigure}
	\begin{subfigure}[b]{\subfigwidth}
		\includetikz{dvcm_gaussDiff_brainstem}
		\caption{brainstem, \ac{DVCM}$_{\mathcal{N}}$ -- \ac{DVCM}$_{\mathcal{S}}$}
		\label{fig:pdvhs:brainstem_n}
	\end{subfigure}
	
	\begin{subfigure}[b]{\subfigwidth}
		\includetikz{dvcm_betaDiff_CTV}
		\caption{\ac{CTV}, \ac{DVCM}$_{\mathcal{B}}$ -- \ac{DVCM}$_{\mathcal{S}}$}
		\label{fig:pdvhs:ctv_b}
	\end{subfigure}
	\begin{subfigure}[b]{\subfigwidth}
		\includetikz{dvcm_betaDiff_brainstem}
		\caption{brainstem, \ac{DVCM}$_{\mathcal{B}}$ -- \ac{DVCM}$_{\mathcal{S}}$}
		\label{fig:pdvhs:brainstem_b}
	\end{subfigure}
	\caption{Evaluation of probabilities and $\alpha$-\acp{DVH} obtained through scenario sampling and analytical computations. (a) and (b) show empirical \acp{DVCM} (indexed with $\mathcal{S}$) obtained from the \num{5000} dose scenarios, where the color indicates the local value of the \ac{CDF} of the respective \ac{DVH}-point. Corresponding $\alpha$-\acp{DVH} were derived, based on sampling ($\mathcal{S}$) and on \textcite{Henriquez2010} ($\mathcal{H}$). 
		For (c) and (d) the difference of \ac{DVCM}$_{\mathcal{N}}$, a \ac{DVCM} constructed from the normal parameterization, to \ac{DVCM}$_{\mathcal{S}}$ has been evaluated with corresponding $\alpha$-\acp{DVH} obtained from the normal quantile function. (e,f) provide a similar analysis using \ac{DVCM}$_{\mathcal{B}}$ and $\alpha$-\acp{DVH} ($\mathcal{B}$) with the beta parameterization.}
	\label{fig:pdvhs}
\end{figure}

In \cref{fig:pdvhs:ctv_n,fig:pdvhs:brainstem_n}, the probabilities within the empirical \acp{DVCM} show large differences when compared to \acp{DVCM} computed with the \ac{CDF} from the Gaussian approximation, especially near full volume coverage and near zero volume coverage, where the approximated \ac{CDF} exhibits differences of up to \num{+-0.2}.
This is to be expected due to the infinite support of the normal distribution, and therefore the agreement is much better with the beta approximation in \cref{fig:pdvhs:ctv_b,fig:pdvhs:brainstem_b} (overall the deviation is more than halved compared to the normal approximation). This transfers to the computation of $\alpha$-\acp{DVH} based on the quantile function of the beta distribution, similarly showing better agreement than with the assumption of normally distributed \ac{DVH}-points. Overall, our explicit parametrization and evaluation of quantile functions of either normal or especially beta distributions 
\added{outperforms the previously published method\autocite{Henriquez2010} (compare \cref{sec:previous_works_pdvh}).}
\deleted{is superior to the method from \textcite{Henriquez2010}.}

\section{Discussion}
\label{sec:discussion}
The very core of this work is the description of an analytical model that can compute statistical moments of \ac{DVH}-points for arbitrary probability distributions over the dose distribution. The only requirement is that the dose distribution follows/obeys a probability distribution function where the marginal \acp{CDF} can be evaluated. 
Hence, we provide a mathematically exact formulation of the moments which eliminates statistical uncertainty (in particular in combination with \ac{APM}-based uncertainty quantification \autocite{Bangert2013,Wahl2018}); if the probability distribution over dose is known, the respective moments can be exactly computed. 

\added{In our initial analysis, a multivariate normal distribution was assumed based on expected dose and its covariance obtained from a probabilistic dose calculation employing \ac{APM}, which we compared to a sampling benchmark. Most of the analytically computed expected and standard deviations of the investigated \acp{DVH} lied within \SI{+-1.5}{\percent} difference in volume from the sampling benchmark. Evaluating the same model for a larger number of fractions showed that especially for the \acp{OAR}, more values cluster around the small differences. 

Those differences between analytical model and sampling benchmark mainly stem from the assumed probability distribution over the dose distribution. It is clear that the multivariate normal assumption for dose uncertainty used throughout this work does not reflect every detail of reality perfectly. This is directly obvious because dose cannot be negative while the normal distribution has infinite support. But also deformations and anatomical variations, the non-linearity of dose deposition, and the so far not captured complexity of input uncertainty suggest that \enquote{real} dose uncertainty will most likely manifest as non-Gaussian probability distributions differing across heavily interdependent voxels. For parametrizing such models, an appropriate family of probability distributions that reasonably models variation in voxel doses would have to be selected for which correlations could then, for example, be modeled with copulas. While finding a suitable probabilistic model accurately describing dose distributions under uncertainty is beyond the scope of this work, our method would also be compatible with such advanced models, since it only requires that some form of multivariate \ac{CDF} can be computed. This includes non-Gaussian analytical as well as empirical \acp{CDF}. In doing so our model has the advantage to preserve the mathematical connection between dose \ac{DVH}-variation which would not be accessible when sampling \ac{DVH}-scenarios themselves. 

Despite the inaccuracy inherent to our choice of a multivariate normal distribution, it served as good initial proof of concept for the presented method. The multivariate normal parametrization allowed straightforward resampling of dose scenarios to demonstrate the exactness of the closed-form solution.  The validation with results from sampled dose cubes and \acp{DVH} on the three patient cases showed that despite using this physically flawed model, we still obtain reasonable results, especially for the prostate and paraspinal case. The larger deviations in the intracranial case may be attributed to the smaller \acp{VOI} compared to the other two cases. The above discussed observation that fractionation seems to decrease the difference in most voxels might be attributed to a more Gaussian shaped probability distribution of dose uncertainty in the context of fractionation \parencite[as also indicated within][]{Lowe2017}.}

\deleted{It is clear that the multivariate normal assumption for dose uncertainty does not reflect reality to \mbox{\SI{100}{\percent}}, but served as a good initial proof of concept for the presented method. The validation with results from sampled dose cubes and \acp{DVH} on the three patient cases showed that despite using this physically flawed model, we still obtain reasonable results, especially for the prostate and paraspinal case. The larger deviations in the intracranial case may be attributed to the smaller \acp{VOI} compared to the other two cases.}

\deleted{Most of the analytically computed expected and standard deviations of \ac{DVH} lied within \mbox{\SI{+-1.5}{\percent}} difference in volume from the sampling benchmark. Evaluating the same model for a larger number of fractions showed that especially for the \acp{OAR}, more values cluster around the small differences. A reason for this could be that fractionation induces an additional component leading to a more Gaussian shaped distribution of dose uncertainty {\parencite[as also shown within][]{Lowe2017}}. It would be further possible to refine the analytical model by using other, more appropriate distributions, possibly in combination with copulas to model the correlation for arbitrary marginal distributions.}

\replaced{Our model}{Since the method solely} describes moments of the \ac{DVH}-points' probability distributions and \added{not} the probability distribution over the full \acp{DVH}\replaced{. Therefore}{,} it is not directly possible to exactly compute confidences on actual realizations of the full \ac{DVH}. It is, however, possible to use the computed moments to parameterize distributions over \deleted{single} \ac{DVH} points. \replaced{In our evaluations based on}{Since we evaluated} expected \ac{DVH}-points and their standard deviation, usage of normal distributions was a first obvious choice. To some extent it may be surprising that this yielded acceptable results within few volume percent difference to the sampling benchmark: \deleted{Since} \replaced{T}{t}he choice of a normal distribution (with infinite support) is in th\replaced{is}{e} case of a volume fraction, which can take only values in the interval $[0,1]$, \replaced{similarly}{at least as} physically unreasonable as in the case of the dose distribution. A more plausible (concerning the support interval) parameterization was found using a beta distribution. However, both distributions do not represent a mathematically exact model. Interpreting the calculation of a \ac{DVH}-point as a series of Bernoulli trials \parencite[similar to ][see \cref{sec:previous_works_expDVH}]{Henriquez2008}, suggests parameterization with a Binomial distribution in the case of independently and identically distributed individual voxel doses. Still, assuming independence of voxels would not be realistic. More suitable correlated binomial models \cite{Hisakado2006} make specific assumptions and are computationally demanding, rendering them not applicable for our purpose. Nevertheless, assuming a beta distribution (or even a normal distribution), which is parameterized by mean and standard deviation for the \ac{DVH}-point, could facilitate uncertainty propagation through models that build on \acp{DVH} itself, \eg in deriving biologically effective dose \autocite{Wheldon1998}, or refining the statistical models for optimization purposes \autocite{Sobotta2010}.

In comparison to the previous \replaced{works\autocite{Henriquez2008,Henriquez2008a,Henriquez2010}}{works of \textcite{Henriquez2008,Henriquez2008a,Henriquez2010}}, our \replaced{methodology directly reproduced their model}{model could reproduce their result} for the expectation value of \ac{DVH}-points \autocite{Henriquez2008}. \replaced{Our approach using integration directly generalizes to higher moments like covariance, while their model only yielded upper bounds on the variance because correlations between voxels were not explicitly included.}{While our model at the same time generalizes to higher moments, they did not attempt to compute higher moments due to a simplified dose uncertainty lacking explicit modeling of correlation between voxels.}
Yet, despite the lack of an uncertainty model considering covariance in dose, they \replaced{formulated}{attempted to derive} confidence bounds, i.\,e., $\alpha$-\acp{DVH} \autocite{Henriquez2010}. Those were, however, in stark disagreement with the sampling benchmark (and thence our model). \added{This may mainly be attributed to neglecting the contribution of (in)dependencies between voxels in the derivations.}
\replaced{The}{In fact, the} difference of \cref{eq:pdvh_hc} to \replaced{a \enquote{true}}{an actual} confidence of dose over coverage of a volume fraction may be shown by the following gedankenexperiment: Let us assume all $V$ voxels in a \ac{VOI} $v$ are independently normally distributed with a mean value of $\hat{d}$ and identical variances, and therefore exhibiting $F_{d_i}(\hat{d}) = \num{0.5}$ in all voxels $i \in v$. 
In this setup, \replaced{one can see}{it is trivial to see} that the median \ac{DVH}-point at $\hat{d}$ takes the value $\num{0.5}$. Yet, plugging this into \cref{eq:pdvh_hc} with again $\alpha=\num{0.5}$ yields $\alpha\text{-DVH}^{\text{HC}}(\hat{d}) = \frac{1}{\numvox}\cdot\numvox\cdot\Theta(1-0.5-0.5) = 1 \neq \num{0.5}$ (depending on the definition of the Heaviside-step). Furthermore, in this case \cref{eq:pdvh_hc} exhibits a sharp \enquote{step} around $\hat{d}$: For smaller doses $\hat{d} - \epsilon$, the argument of the step function in \cref{eq:pdvh_hc} becomes negative and therefore resulting in $\alpha\text{-DVH}(\hat{d}-\epsilon) = 1$, while for larger doses it becomes $\alpha\text{-DVH}(\hat{d}-\epsilon) = 0$. 
\replaced{For the assumed independently distributed model, however, a smooth decrease of the median \ac{DVH} around $\hat{d}$ is expected, whereas the step would be correctly reproduced when assuming perfectly correlated voxels.}{It is clear that this result is absolutely unreasonable for the assumed independently distributed model, where a smooth decrease of the median \ac{DVH} around $\hat{d}$ is expected, whereas when assuming perfectly correlated voxels, a similar step in the median \ac{DVH} will form. Therefore, we suggest to not interpret their result as an $\alpha$-\ac{DVH} in its classical sense according to literature.} 
\deleted{Consequently, their model is clearly not representing an $\alpha$-\ac{DVH} in its classical sense and therefore gives misleading results, instead} 
\added{Instead} we suggest to interpret their result as the \enquote{fraction of voxels whose probability of exceeding $\hat{d}$ \emph{independently from each other} is larger than $1-\alpha$}. Such a quantity, however, does not have a palpable clinical interpretation.

The applicability of a method that propagates uncertainty from dose to \acp{DVH} in terms of treatment planning might not be directly obvious. As discussed in \cref{sec:introduction}, uncertainty quantification usually relies on sampled (stochastic approach) or selected (worst-case approach) dose scenarios, which can directly be used to obtain similar uncertainty information about the \ac{DVH}. However, the analytical probabilistic method presented here provides a closed-form, continuous relationship between dose uncertainty and \ac{DVH}-uncertainty, which can be useful in treatment planning\replaced{.}{, especially for optimization purposes with probabilistic constraints.}

\replaced{For optimization purposes with probabilistic constraints}{There}, the method facilitates the use of continuous differentiable functions, which \enquote{fill the gap} between empirical samples, and could possibly enable exact definition of the \enquote{allowed} probability that certain clinical constraints are failing. 
\added{There, using the parameterization of the probability distribution over a \ac{DVH}-point, we now have functional access to the probability that a specific dose level is covered with a certain probability via the $\alpha$-\ac{DVH} and its analytical dependence on mean and covariance of the dose (instead of nominal dose as for conventional non-probabilisitc objectives).\autocite{Wahl2018a,Wahl2019a} Using this information for planning would extend widely used margin principles\autocite{vanHerk2000,vanHerk2004}, which are themselves defined based on demanding tumor coverage with a desired probability for a given dose level.	Compared to common robust approaches where the worst case is dynamically found from a discrete pre-defined uncertainty subset in the input variables, an analytical confidence constraint based on $\alpha$-\acp{DVH} captures the full uncertainty space and allows for  definition of the desired robustness on the relevant endpoint. For example, \SI{90}{\percent} probability to cover a tumor with the \SI{95}{\percent} of the prescribed dose could reproduce common margin assumptions also for anatomically complex cases where margins break down, while at the same time other probabilistic or nominal endpoints could be optimized.  Since these objectives (or also constraint functions) continue to be functions of mean and covariance of dose,  optimization would remain possible using conventional (quasi-)Newton optimizers. This, however, comes at the cost of deriving the arguably complex gradients (w.r.t.\ dose mean \& covariance), which could be mitigated by using automatic differentiation \autocite{Griewank2008}, accompanied by a computationally challenging, yet \enquote{embarrassingly parallel} (co)variance computation.\autocite{Wahl2018a,Wahl2019a} Further, the role of model inaccuracy in such a probabilistic optimization approach would need to be further investigated---\textcite{Wieser2020}, however, suggest that the assumption of normal distributions in probabilistic optimization still enables mitigation of dose uncertainties even if the \enquote{real} distribution differs substantially. A full analysis of computational performance and resulting dose distributions of such an optimization approach would exceed this manuscript and may be part of a future study.}
 
Further, the probabilistic model only requires the dose's probability distribution and is independent of the method used to obtain it. This allows its general application, \eg,  in \added{(retrospective) plan}\deleted{retrospective} analyses or in combination with sampling-based stochastic optimization approaches. \added{For plan analysis and comparison, the closed-form connection of dose and \ac{DVH}-uncertainty could be used to observe how higher dose uncertainty (simulated by scaling covariance in desired voxels, for example) would impact the \ac{DVH} and therefore target coverage. When relying on sample statistics for the dose instead of a probabilstic dose calculation, the parameterizations can be useful to extrapolate to areas not covered by samples. Additionally, the method could also be used in scenarios where computation of uncertainties statistically is no longer possible (\eg when analysing an older cohort) based on assumptions over the dose probability distributions while still retaining the mathematical link between dose and \ac{DVH} probabilities, which may be useful for machine learning applications investigating plan uncertainties.}

And last but not least, the generral concept does not only apply to the computation of \ac{DVH} points. As already indicated by \textcite{Wahl2018a}, the concept may---given \deleted{some} mathematical efforts---be extended to other plan quality metrics as mean dose or \ac{EUD}, or distinct treatment planning objective and constraint functions.

\section{Conclusion}
We presented a method to calculate moments of the probability distribution over \ac{DVH}-points given a known probability distribution over the dose distribution. The resulting analytical model corrects and generalizes previous attempts and can be readily combined with every method able to quantify a probability distribution over dose of either empirical or probabilistic nature. 

We successfully benchmarked the model against excessive sampling (with and without fractionation), proving mathematical correctness and good agreement even with unrealistic but common assumptions for probability distributions within the computational pipeline. The methodology can serve as blueprint for future models on other \acp{QI} and can provide a generalizable framework for confidence-constrained probabilistic treatment plan optimization.

\printbibliography

@article{Bangert2013,
  title = {Analytical probabilistic modeling for radiation therapy treatment planning.},
  author = {Bangert, Mark and Hennig, Philipp and Oelfke, Uwe},
  date = {2013},
  journaltitle = {Physics in Medicine and Biology},
  shortjournal = {Phys Med Biol},
  volume = {58},
  pages = {5401--5419},
  issn = {1361-6560},
  doi = {10.1088/0031-9155/58/16/5401},
  eprint = {23877218},
  eprinttype = {pmid},
  number = {16}
}

@article{Bohoslavsky2013,
  title = {Probabilistic objective functions for margin-less {{IMRT}} planning},
  author = {Bohoslavsky, Román and Witte, Marnix G and Janssen, Tomas M and family=Herk, given=Marcel, prefix=van, useprefix=true},
  date = {2013},
  journaltitle = {Physics in Medicine and Biology},
  shortjournal = {Phys Med Biol},
  volume = {58},
  pages = {3563--3580},
  issn = {0031-9155},
  doi = {10.1088/0031-9155/58/11/3563},
  url = {http://stacks.iop.org/0031-9155/58/i=11/a=3563?key=crossref.79938efd7333300ef1b285f9f13502c7},
  eprint = {23640114},
  eprinttype = {pmid},
  number = {11}
}

@article{Casiraghi2013a,
  title = {Advantages and limitations of the 'worst case scenario' approach in {{IMPT}} treatment planning.},
  author = {Casiraghi, M and Albertini, F and Lomax, Antony John},
  date = {2013-03-07},
  journaltitle = {Physics in Medicine and Biology},
  shortjournal = {Phys Med Biol},
  volume = {58},
  pages = {1323--1339},
  publisher = {{IOP Publishing}},
  issn = {1361-6560},
  doi = {10.1088/0031-9155/58/5/1323},
  eprint = {23391569},
  eprinttype = {pmid},
  number = {5}
}

@article{Fredriksson2011,
  title = {Minimax optimization for handling range and setup uncertainties in proton therapy},
  author = {Fredriksson, Albin and Forsgren, Anders and Hårdemark, Björn},
  date = {2011},
  journaltitle = {Medical Physics},
  shortjournal = {Med Phys},
  volume = {38},
  pages = {1672--1684},
  publisher = {{American Association of Physicists in Medicine}},
  issn = {00942405},
  doi = {10.1118/1.3556559},
  url = {http://scitation.aip.org/content/aapm/journal/medphys/38/3/10.1118/1.3556559},
  urldate = {2016-07-25},
  eprint = {21520880},
  eprinttype = {pmid},
  number = {3}
}

@article{Fredriksson2012,
  title = {A characterization of robust radiation therapy treatment planning methods—from expected value to worst case optimization},
  author = {Fredriksson, Albin},
  date = {2012-07-31},
  journaltitle = {Medical Physics},
  shortjournal = {Med Phys},
  volume = {39},
  pages = {5169--5181},
  publisher = {{American Association of Physicists in Medicine}},
  issn = {00942405},
  doi = {10.1118/1.4737113},
  url = {http://scitation.aip.org/content/aapm/journal/medphys/39/8/10.1118/1.4737113},
  urldate = {2017-01-13},
  eprint = {22894442},
  eprinttype = {pmid},
  number = {8}
}

@article{Fredriksson2016,
  title = {The scenario-based generalization of radiation therapy margins},
  author = {Fredriksson, Albin and Bokrantz, Rasmus},
  date = {2016-03-07},
  journaltitle = {Physics in Medicine and Biology},
  shortjournal = {Phys Med Biol},
  volume = {61},
  pages = {2067--2082},
  publisher = {{IOP Publishing}},
  issn = {0031-9155},
  doi = {10.1088/0031-9155/61/5/2067},
  url = {http://stacks.iop.org/0031-9155/61/i=5/a=2067?key=crossref.a79f05a7a0b46cfc0673e2002e5c1130},
  urldate = {2016-07-22},
  archivePrefix = {arXiv},
  eprint = {26895381},
  eprinttype = {pmid},
  number = {5}
}

@article{Genz2004,
  title = {Numerical computation of rectangular bivariate and trivariate normal and t probabilities},
  author = {Genz, Alan},
  date = {2004-08},
  journaltitle = {Statistics and Computing},
  shortjournal = {Stat Comput},
  volume = {14},
  pages = {251--260},
  issn = {0960-3174},
  doi = {10.1023/B:STCO.0000035304.20635.31},
  url = {http://link.springer.com/10.1023/B:STCO.0000035304.20635.31},
  number = {3}
}

@article{Gordon2009,
  title = {Coverage-based treatment planning: optimizing the {{IMRT PTV}} to meet a {{CTV}} coverage criterion.},
  author = {Gordon, J J and Siebers, J V},
  date = {2009},
  journaltitle = {Medical Physics},
  shortjournal = {Med Phys},
  volume = {36},
  pages = {961--973},
  issn = {00942405},
  doi = {10.1118/1.3075772},
  eprint = {19378757},
  eprinttype = {pmid},
  number = {3}
}

@article{Gordon2010,
  title = {Coverage optimized planning: probabilistic treatment planning based on dose coverage histogram criteria.},
  author = {Gordon, J J and Sayah, N and Weiss, E and Siebers, J V},
  date = {2010},
  journaltitle = {Medical Physics},
  shortjournal = {Med Phys},
  volume = {37},
  pages = {550--563},
  issn = {00942405},
  doi = {10.1118/1.3273063},
  eprint = {20229863},
  eprinttype = {pmid},
  number = {2}
}

@book{Griewank2008,
  title = {Evaluating {{Derivatives}}},
  author = {Griewank, Andreas and Walther, Andrea},
  date = {2008-01-01},
  publisher = {{Society for Industrial and Applied Mathematics}},
  doi = {10.1137/1.9780898717761},
  url = {https://epubs.siam.org/doi/book/10.1137/1.9780898717761},
  urldate = {2020-06-10},
  isbn = {978-0-89871-659-7},
  pagetotal = {448},
  series = {Other {{Titles}} in {{Applied Mathematics}}}
}

@article{Henriquez2008,
  title = {A {{Novel Method}} for the {{Evaluation}} of {{Uncertainty}} in {{Dose}}-{{Volume Histogram Computation}}},
  author = {Cutanda Henríquez, Francisco and Vargas Castrillón, Silvia},
  date = {2008},
  journaltitle = {International Journal of Radiation Oncology*Biology*Physics},
  shortjournal = {Int J Radiat Oncol Biol Phys},
  volume = {70},
  pages = {1263--1271},
  issn = {03603016},
  doi = {10.1016/j.ijrobp.2007.11.038},
  eprint = {18313532},
  eprinttype = {pmid},
  number = {4}
}

@article{Henriquez2008a,
  title = {The effect of the different uncertainty models in dose expected volume histogram computation},
  author = {Cutanda Henríquez, Francisco and Vargas Castrillón, Silvia},
  date = {2008},
  journaltitle = {Australasian Physical and Engineering Sciences in Medicine},
  shortjournal = {Australas Phys Eng Sci Med},
  volume = {31},
  pages = {196--202},
  issn = {01589938},
  number = {3}
}

@article{Henriquez2010,
  title = {Confidence intervals in dose volume histogram computation},
  author = {Cutanda Henríquez, Francisco and Vargas Castrillón, Silvia},
  date = {2010-03-15},
  journaltitle = {Medical Physics},
  shortjournal = {Med Phys},
  volume = {37},
  pages = {1545--1553},
  issn = {00942405},
  doi = {10.1118/1.3355888},
  url = {http://doi.wiley.com/10.1118/1.3355888},
  eprint = {20443475},
  eprinttype = {pmid},
  number = {4}
}

@article{Hisakado2006,
  title = {Correlated {{Binomial Models}} and {{Correlation Structures}}},
  author = {Hisakado, M. and Kitsukawa, K. and Mori, S.},
  date = {2006-12-15},
  journaltitle = {Journal of Physics A: Mathematical and General},
  shortjournal = {J Phys A Math Gen},
  volume = {39},
  pages = {15365--15378},
  issn = {0305-4470, 1361-6447},
  doi = {10.1088/0305-4470/39/50/005},
  archivePrefix = {arXiv},
  eprint = {physics/0605189},
  eprinttype = {arxiv},
  number = {50}
}

@article{Kraan2013,
  title = {Dose uncertainties in {{IMPT}} for oropharyngeal cancer in the presence of anatomical, range, and setup errors},
  author = {Kraan, Aafke C. and Van De Water, Steven and Teguh, David N. and Al-Mamgani, Abrahim and Madden, Tom and Kooy, Hanne M. and Heijmen, Ben J M and Hoogeman, Mischa S.},
  date = {2013},
  journaltitle = {International Journal of Radiation Oncology*Biology*Physics},
  shortjournal = {Int J Radiat Oncol Biol Phys},
  volume = {87},
  pages = {888--896},
  publisher = {{Elsevier Inc.}},
  issn = {03603016},
  doi = {10.1016/j.ijrobp.2013.09.014},
  url = {http://dx.doi.org/10.1016/j.ijrobp.2013.09.014},
  eprint = {24351409},
  eprinttype = {pmid},
  number = {5}
}

@article{Liu2012,
  title = {Robust optimization of intensity modulated proton therapy},
  author = {Liu, Wei and Zhang, Xiaodong and Li, Yupeng and Mohan, Radhe},
  date = {2012-02},
  journaltitle = {Medical Physics},
  shortjournal = {Med Phys},
  volume = {39},
  pages = {1079--1091},
  publisher = {{American Association of Physicists in Medicine}},
  issn = {0094-2405},
  doi = {10.1118/1.3679340},
  url = {http://scitation.aip.org/content/aapm/journal/medphys/39/2/10.1118/1.3679340},
  urldate = {2016-07-25},
  eprint = {22320818},
  eprinttype = {pmid},
  number = {2}
}

@article{Lomax2008,
  title = {Intensity modulated proton therapy and its sensitivity to treatment uncertainties 1: the potential effects of calculational uncertainties.},
  author = {Lomax, Antony John},
  date = {2008},
  journaltitle = {Physics in Medicine and Biology},
  shortjournal = {Phys Med Biol},
  volume = {53},
  pages = {1027--1042},
  issn = {0031-9155},
  doi = {10.1088/0031-9155/53/4/015},
  eprint = {18263956},
  eprinttype = {pmid},
  number = {4}
}

@article{Lomax2008b,
  title = {Intensity modulated proton therapy and its sensitivity to treatment uncertainties 2: the potential effects of inter-fraction and inter-field motions},
  author = {Lomax, A J},
  date = {2008-02-21},
  journaltitle = {Physics in Medicine and Biology},
  shortjournal = {Phys Med Biol},
  volume = {53},
  pages = {1043--1056},
  issn = {0031-9155},
  doi = {10.1088/0031-9155/53/4/015},
  url = {http://stacks.iop.org/0031-9155/53/i=4/a=015?key=crossref.49a5dabe348f26645479f8d68643b59d},
  urldate = {2018-04-01},
  number = {4}
}

@article{Lowe2016,
  title = {Incorporating the effect of fractionation in the evaluation of proton plan robustness to setup errors},
  author = {Lowe, Matthew and Albertini, Francesca and Aitkenhead, Adam and Lomax, Antony John and MacKay, Ranald I},
  date = {2016-01-07},
  journaltitle = {Physics in Medicine and Biology},
  shortjournal = {Phys Med Biol},
  volume = {61},
  pages = {413--429},
  publisher = {{IOP Publishing}},
  issn = {0031-9155},
  doi = {10.1088/0031-9155/61/1/413},
  eprint = {26675133},
  eprinttype = {pmid},
  number = {1}
}

@article{Lowe2017,
  title = {A robust optimisation approach accounting for the effect of fractionation on setup uncertainties},
  author = {Lowe, Matthew and Aitkenhead, Adam and Albertini, Francesca and Lomax, Antony J and MacKay, Ranald I},
  date = {2017-10-04},
  journaltitle = {Physics in Medicine and Biology},
  shortjournal = {Phys Med Biol},
  volume = {62},
  pages = {8178--8196},
  publisher = {{IOP Publishing}},
  issn = {1361-6560},
  doi = {10.1088/1361-6560/aa8c58},
  url = {http://stacks.iop.org/0031-9155/62/i=20/a=8178?key=crossref.65fca1cc1e4869e819bdda018cd1c9f2},
  urldate = {2018-03-09},
  number = {20}
}

@article{McGowan2015,
  title = {Defining robustness protocols: a method to include and evaluate robustness in clinical plans.},
  author = {McGowan, S E and Albertini, F and Thomas, S J and Lomax, A J},
  date = {2015-04-07},
  journaltitle = {Physics in Medicine and Biology},
  shortjournal = {Phys Med Biol},
  volume = {60},
  pages = {2671--2684},
  publisher = {{IOP Publishing}},
  issn = {1361-6560},
  doi = {10.1088/0031-9155/60/7/2671},
  url = {http://stacks.iop.org/0031-9155/60/i=7/a=2671?key=crossref.092f8f208f90dc19331c51799573c57f},
  urldate = {2015-03-17},
  eprint = {25768095},
  eprinttype = {pmid},
  langid = {english},
  number = {7}
}

@article{Mescher2017,
  title = {Coverage-based constraints for {{IMRT}} optimization},
  author = {Mescher, H and Ulrich, S and Bangert, M},
  date = {2017-09-05},
  journaltitle = {Physics in Medicine and Biology},
  shortjournal = {Phys Med Biol},
  volume = {62},
  pages = {N460-N473},
  publisher = {{IOP Publishing}},
  issn = {1361-6560},
  doi = {10.1088/1361-6560/aa8132},
  url = {http://stacks.iop.org/0031-9155/62/i=18/a=N460?key=crossref.f05e3801e9bf4cfba9eb57e115758556},
  urldate = {2018-04-13},
  number = {18}
}

@article{Moore2009,
  title = {Comparisons of treatment optimization directly incorporating random patient setup uncertainty with a margin-based approach.},
  author = {Moore, Joseph A and Gordon, John J and Anscher, Mitchell S and Siebers, Jeffrey V},
  date = {2009},
  journaltitle = {Medical Physics},
  shortjournal = {Med Phys},
  volume = {36},
  pages = {3880--3890},
  issn = {0094-2405},
  doi = {10.1118/1.3176940},
  eprint = {19810460},
  eprinttype = {pmid},
  number = {9}
}

@article{Nadarajah2017,
  title = {A new bivariate beta distribution},
  author = {Nadarajah, Saralees and Shih, Shou Hsing and Nagar, Daya K.},
  date = {2017-03-04},
  journaltitle = {Statistics},
  shortjournal = {Statistics},
  volume = {51},
  pages = {455--474},
  publisher = {{Taylor \& Francis}},
  issn = {0233-1888},
  doi = {10.1080/02331888.2016.1240681},
  url = {https://www.tandfonline.com/doi/full/10.1080/02331888.2016.1240681},
  urldate = {2018-07-24},
  number = {2}
}

@article{Olkin2015,
  title = {Constructions for a bivariate beta distribution},
  author = {Olkin, Ingram and Trikalinos, Thomas A.},
  date = {2015-01-01},
  journaltitle = {Statistics \& Probability Letters},
  shortjournal = {Stat Probab Lett},
  volume = {96},
  pages = {54--60},
  publisher = {{North-Holland}},
  issn = {01677152},
  doi = {10.1016/j.spl.2014.09.013},
  url = {https://www.sciencedirect.com/science/article/pii/S0167715214003241},
  urldate = {2018-06-29},
  archivePrefix = {arXiv},
  eprint = {1406.5881},
  eprinttype = {arxiv}
}

@article{Park2013,
  title = {Statistical assessment of proton treatment plans under setup and range uncertainties},
  author = {Park, Peter C. and Cheung, Joey P. and Zhu, X. Ronald and Lee, Andrew K. and Sahoo, Narayan and Tucker, Susan L. and Liu, Wei and Li, Heng and Mohan, Radhe and Court, Laurence E. and Dong, Lei},
  date = {2013},
  journaltitle = {International Journal of Radiation Oncology*Biology*Physics},
  shortjournal = {Int J Radiat Oncol Biol Phys},
  volume = {86},
  pages = {1007--1013},
  publisher = {{Elsevier Inc.}},
  issn = {03603016},
  doi = {10.1016/j.ijrobp.2013.04.009},
  url = {http://dx.doi.org/10.1016/j.ijrobp.2013.04.009},
  eprint = {23688812},
  eprinttype = {pmid},
  number = {5}
}

@article{Perko2016,
  title = {Fast and accurate sensitivity analysis of {{IMPT}} treatment plans using {{Polynomial Chaos Expansion}}.},
  author = {Perkó, Zoltán and family=Voort, given=Sebastian R, prefix=van der, useprefix=true and family=Water, given=Steven, prefix=van de, useprefix=true and Hartman, Charlotte M H and Hoogeman, Mischa and Lathouwers, Danny},
  date = {2016-06-21},
  journaltitle = {Physics in Medicine and Biology},
  shortjournal = {Phys Med Biol},
  volume = {61},
  pages = {4646--4664},
  publisher = {{IOP Publishing}},
  issn = {1361-6560},
  doi = {10.1088/0031-9155/61/12/4646},
  url = {http://stacks.iop.org/0031-9155/61/i=12/a=4646?key=crossref.0937074c86bf618a62c04c86e4929f71},
  urldate = {2016-07-29},
  eprint = {27227661},
  eprinttype = {pmid},
  number = {12}
}

@article{Pflugfelder2008,
  title = {Worst case optimization: a method to account for uncertainties in the optimization of intensity modulated proton therapy.},
  author = {Pflugfelder, Daniel and Wilkens, Jan Jakob and Oelfke, Uwe},
  date = {2008-03-21},
  journaltitle = {Physics in Medicine and Biology},
  shortjournal = {Phys Med Biol},
  volume = {53},
  pages = {1689--700},
  publisher = {{IOP Publishing}},
  issn = {0031-9155},
  doi = {10.1088/0031-9155/53/6/013},
  url = {http://stacks.iop.org/0031-9155/53/i=6/a=013?key=crossref.f26e33d9f0cb93a513fbd6d79e3981f0},
  urldate = {2016-07-25},
  eprint = {18367797},
  eprinttype = {pmid},
  number = {6}
}

@article{Sobotta2010,
  title = {Robust optimization based upon statistical theory},
  author = {Sobotta, B. and Söhn, M. and Alber, M.},
  date = {2010-07-13},
  journaltitle = {Medical Physics},
  shortjournal = {Med Phys},
  volume = {37},
  pages = {4019--4028},
  publisher = {{American Association of Physicists in Medicine}},
  issn = {00942405},
  doi = {10.1118/1.3457333},
  url = {http://doi.wiley.com/10.1118/1.3457333},
  urldate = {2017-02-03},
  number = {8}
}

@article{Sobotta2012,
  title = {Accelerated evaluation of the robustness of treatment plans against geometric uncertainties by {{Gaussian}} processes},
  author = {Sobotta, B and Söhn, M and Alber, M},
  date = {2012-12-07},
  journaltitle = {Physics in Medicine and Biology},
  shortjournal = {Phys Med Biol},
  volume = {57},
  pages = {8023--8039},
  publisher = {{IOP Publishing}},
  issn = {0031-9155},
  doi = {10.1088/0031-9155/57/23/8023},
  url = {http://stacks.iop.org/0031-9155/57/i=23/a=8023?key=crossref.bcc5b813db7008ce072e53460d2f79e6},
  urldate = {2016-11-01},
  number = {23}
}

@article{Unkelbach2007,
  title = {Accounting for range uncertainties in the optimization of intensity modulated proton therapy.},
  author = {Unkelbach, Jan and Chan, Timothy C Y and Bortfeld, Thomas},
  date = {2007-05-21},
  journaltitle = {Physics in Medicine and Biology},
  shortjournal = {Phys Med Biol},
  volume = {52},
  pages = {2755--2773},
  issn = {0031-9155},
  doi = {10.1088/0031-9155/52/10/009},
  url = {http://stacks.iop.org/0031-9155/52/i=10/a=009},
  urldate = {2015-03-25},
  eprint = {17473350},
  eprinttype = {pmid},
  number = {10}
}

@article{Unkelbach2009,
  title = {Reducing the sensitivity of {{IMPT}} treatment plans to setup errors and range uncertainties via probabilistic treatment planning.},
  author = {Unkelbach, Jan and Bortfeld, Thomas and Martin, Benjamin C. and Soukup, Martin},
  date = {2009-12-12},
  journaltitle = {Medical Physics},
  shortjournal = {Med Phys},
  volume = {36},
  pages = {149--163},
  publisher = {{American Association of Physicists in Medicine}},
  issn = {00942405},
  doi = {10.1118/1.3021139},
  url = {http://scitation.aip.org/content/aapm/journal/medphys/36/1/10.1118/1.3021139},
  urldate = {2015-03-25},
  eprint = {19235384},
  eprinttype = {pmid},
  number = {2009}
}

@article{Unkelbach2018,
  title = {Robust radiotherapy planning},
  author = {Unkelbach, Jan and Alber, Markus and Bangert, Mark and Bokrantz, Rasmus and Chan, Timothy C Y and Deasy, Joseph O and Fredriksson, Albin and Gorissen, Bram L and family=Herk, given=Marcel, prefix=van, useprefix=true and Liu, Wei and Mahmoudzadeh, Houra and Nohadani, Omid and Siebers, Jeffrey V and Witte, Marnix and Xu, Huijun},
  date = {2018-11-12},
  journaltitle = {Physics in Medicine and Biology},
  shortjournal = {Phys Med Biol},
  volume = {63},
  pages = {22TR02},
  publisher = {{IOP Publishing}},
  issn = {1361-6560},
  doi = {10.1088/1361-6560/aae659},
  url = {http://stacks.iop.org/0031-9155/63/i=22/a=22TR02?key=crossref.f8f50d2ac0a867a3fa115d936f139c87},
  urldate = {2019-05-20},
  number = {22}
}

@article{vanHerk2000,
  title = {The probability of correct target dosage: dose-population histograms for deriving treatment margins in radiotherapy},
  author = {family=Herk, given=Marcel, prefix=van, useprefix=true and Remeijer, Peter and Rasch, Coen and Lebesque, Joos V.},
  date = {2000},
  journaltitle = {International Journal of Radiation Oncology*Biology*Physics},
  shortjournal = {Int J Radiat Oncol Biol Phys},
  volume = {47},
  pages = {1121--1135},
  url = {http://www.sciencedirect.com/science/article/pii/S0360301600005186},
  urldate = {2014-01-15},
  number = {4}
}

@article{vanHerk2004,
  title = {Errors and margins in radiotherapy},
  author = {family=Herk, given=Marcel, prefix=van, useprefix=true},
  date = {2004-01},
  journaltitle = {Seminars in Radiation Oncology},
  shortjournal = {Semin Radiat Oncol},
  volume = {14},
  pages = {52--64},
  issn = {10534296},
  doi = {10.1053/j.semradonc.2003.10.003},
  url = {http://linkinghub.elsevier.com/retrieve/pii/S1053429603000845},
  urldate = {2017-01-30},
  number = {1}
}

@article{Wahl2017a,
  title = {Efficiency of analytical and sampling-based uncertainty propagation in intensity-modulated proton therapy},
  author = {Wahl, Niklas and Hennig, Philipp and Wieser, Hans-Peter and Bangert, Mark},
  date = {2017-06-26},
  journaltitle = {Physics in Medicine and Biology},
  shortjournal = {Phys Med Biol},
  volume = {62},
  pages = {5790--5807},
  issn = {1361-6560},
  doi = {10.1088/1361-6560/aa6ec5},
  url = {http://stacks.iop.org/0031-9155/62/i=14/a=5790?key=crossref.cba9093365b6707443b63a0770bf9fee},
  number = {14}
}

@article{Wahl2018,
  title = {Analytical incorporation of fractionation effects in probabilistic treatment planning for intensity-modulated proton therapy},
  author = {Wahl, Niklas and Hennig, Philipp and Wieser, Hans-Peter and Bangert, Mark},
  date = {2018-04-27},
  journaltitle = {Medical Physics},
  shortjournal = {Med Phys},
  volume = {45},
  pages = {1317--1328},
  issn = {00942405},
  doi = {10.1002/mp.12775},
  url = {http://doi.wiley.com/10.1002/mp.12775},
  number = {4}
}

@thesis{Wahl2018a,
  title = {Analytical {{Models}} for {{Probabilistic Inverse Treatment Planning}} in {{Intensity}}-modulated {{Proton Therapy}}},
  author = {Wahl, Niklas},
  date = {2018},
  institution = {{Ruprecht-Karls Universität Heidelberg}},
  location = {{Heidelberg}},
  doi = {10.11588/heidok.00025127},
  url = {http://archiv.ub.uni-heidelberg.de/volltextserver/25127/},
  urldate = {2018-08-22},
  type = {Dissertation}
}

@inproceedings{Wahl2019a,
  title = {Confidence constraints for probabilistic radiotherapy treatment planning},
  booktitle = {19th {{International Conference}} on the {{Use}} of {{Computers}} in {{Radiation Therapy}} ({{ICCR}})},
  author = {Wahl, Niklas and Hennig, Philipp and Wieser, Hans-Peter and Bangert, Mark},
  date = {2019}
}

@article{Wheldon1998,
  title = {The linear-quadratic transformation of dose–volume histograms in fractionated radiotherapy},
  author = {Wheldon, Thomas E. and Deehan, Charles and Wheldon, Elizabeth G. and Barrett, Ann},
  date = {1998-03},
  journaltitle = {Radiotherapy and Oncology},
  shortjournal = {Radiother Oncol},
  volume = {46},
  pages = {285--295},
  issn = {01678140},
  doi = {10.1016/S0167-8140(97)00162-X},
  url = {https://linkinghub.elsevier.com/retrieve/pii/S016781409700162X},
  urldate = {2019-09-23},
  langid = {english},
  number = {3}
}

@article{Wieser2017b,
  title = {Analytical probabilistic modeling of {{RBE}}-weighted dose for ion therapy},
  author = {Wieser, Hans-Peter and Hennig, Philipp and Wahl, Niklas and Bangert, Mark},
  date = {2017-10-05},
  journaltitle = {Physics in Medicine and Biology},
  shortjournal = {Phys Med Biol},
  volume = {62},
  pages = {8959--8982},
  publisher = {{IOP Publishing}},
  issn = {0031-9155},
  doi = {10.1088/1361-6560/aa915d},
  url = {http://iopscience.iop.org/article/10.1088/1361-6560/aa915d},
  urldate = {2017-10-05},
  number = {23}
}

@article{Wieser2020,
  title = {Impact of {{Gaussian}} uncertainty assumptions on probabilistic optimization in particle therapy},
  author = {Wieser, Hans-Peter and Karger, Christian P. and Wahl, Niklas and Bangert, Mark},
  date = {2020},
  journaltitle = {Physics in Medicine and Biology},
  shortjournal = {Phys. Med. Biol.},
  issn = {0031-9155},
  doi = {10.1088/1361-6560/ab8d77},
  url = {http://iopscience.iop.org/10.1088/1361-6560/ab8d77},
  urldate = {2020-04-28},
  langid = {english}
}

\ifnum\value{endnote}=0
\else
\theendnotes
\fi

\appendix
\section{Beta distribution}
\label{app:beta}
Suppose a random variable $X$ follows a beta distribution, \ie $X \sim \mathcal{B}(a,b)$ with shape parameters $a$ and $b$.

Within the interval $x \in [0,1]$ (for $a > b > 1$, $x \in (0,1)$ otherwise) its \ac{PDF} $f_X(x)$ is given by
\begin{equation}
	f_X(x) = \frac{1}{\mathrm{B}(a,b)} x^{a-1}(1-x)^{b-1}
\end{equation}
where the normalization $\mathrm{B}(a,b)$ is the beta-function.

The \ac{CDF} 
\begin{equation}
	F_X(x) = I_x(a,b)
\end{equation}
requires evaluation of the regularized incomplete beta-function $I_x$.

Expectation and variance of $X$ are then given by
\begin{align}
	\Exp{X} &= \frac{a}{a + b} \label{eq:app:beta:exp}\,,\\
	\Var{X} &= \frac{ab}{(a + b + 1)(a+b)^2} = \frac{\Exp{X}^2b}{a^2 + ab + a}\label{eq:app:beta:var}\,.
\end{align}

The shape parameters $a$ and $b$ can be inferred from sample statistics, \ie the sample mean $\bar{x}$ and the sample variance $\bar{\sigma}^2$, with the method of moments based on \cref{eq:app:beta:exp,eq:app:beta:var} for $\bar{\sigma}^2 < \bar{x}(1-\bar{x})$:
\begin{align}
	\ensuremath{\hat{a}} &= \bar{x}\left(\frac{\bar{x}(1-\bar{x})}{\bar{\sigma}^2} - 1\right)\label{eq:app:beta:alpha_est}\,,\\
	\ensuremath{\hat{a}} &= (1-\bar{x})\left(\frac{\bar{x}(1-\bar{x})}{\bar{\sigma}^2} - 1\right) = \bar{x}(1-\bar{x})\hat{a}\label{eq:app:beta:beta_est}\,.
\end{align}

\section{Patient Data Information}
\begin{table}[h!]
	\centering%
	\caption{Information on the three patient datasets used for evaluation (similar to \textcite{Wahl2018}).}
	\label{tab:patient_data}
	\footnotesize%
	\sisetup{list-pair-separator={, },list-final-separator={, }}
\begin{tabular}{rccc}
	\toprule%
	patient & intra-cranial & para-spinal & prostate \\
	\midrule%
	beam angles & \SIlist{60;120}{\degree} & \SIlist{135;180;225}{\degree} & \SIlist{90;270}{\degree} \\
	prescribed dose & \SI{60}{\gray} & \SI{60}{\gray} & \SI{70}{\gray} (\SI{76}{\gray}) \\
	beamlet distance & \SI{3}{\milli\metre} &  \SI{4}{\milli\metre} & \SI{5}{\milli\metre}\\
	\#beamlets & 1705 & 13274 & 6803 \\
	resolution &  $(\num{1.2}\times\num{1.2}\times\num{3})\,\si{\milli\metre\cubed}$ & $(\num{3}\times\num{3}\times\num{3})\,\si{\milli\metre\cubed}$ & $(\num{2}\times\num{2}\times\num{3})\,\si{\milli\metre\cubed}$ \\
	setup error \added{[std.\ dev.]} & $(\SI{1}{\milli\metre})^{\text{sys}} + (\SI{2}{\milli\metre})^{\text{rand}}$ & $(\SI{1}{\milli\metre})^{\text{sys}} + (\SI{2}{\milli\metre})^{\text{rand}}$ &
	$(\SI{1}{\milli\metre})^{\text{sys}} + (\SI{3}{\milli\metre})^{\text{rand}}$\\
	range error \added{[std.\ dev.]} & $(\SI{3.5}{\percent})^{\text{sys}} + (\SI{1}{\milli\metre})^{\text{rand}}$ & $(\SI{3.5}{\percent})^{\text{sys}} + (\SI{1}{\milli\metre})^{\text{rand}}$ &
	$(\SI{3.5}{\percent})^{\text{sys}} + (\SI{1}{\milli\metre})^{\text{rand}}$ \\
	\bottomrule
\end{tabular}
%
\end{table}

\section{Generalized model for the \texorpdfstring{$\nu$-th}{Nu-th} moment of a probability distribution over a DVH-point}
\label{app:multi}
Using multi-index notation with the multi-index $\vecbf{\kappa} = \left(\kappa_1, \kappa_2, \ldots, \kappa_{\numvox}\right) \in \mathbb{N}_0^{\numvox}$ and by definition of a multi-indexed Heaviside step $\Theta^{\vecbf{\kappa}}(\vecbf{\tilde{d}} - \hat{d}) = \prod_{i=1}^{n} \Theta(\tilde{d}_{i} - \hat{d})^{\kappa_i}$, one can provide a compact general formula to compute the $\nu$-th non-central moment of the probability distribution of a \ac{DVH}-point, \ie
\begin{equation}
\begin{aligned}
\Exp{\mathrm{DVH}(\hat{d};\vecbf{d})^\nu} &=  \int_{\mathbb{R}^{n}}  \frac{1}{\numvox^\nu}\sum_{|\vecbf{\kappa}| = \nu} \begin{pmatrix} \nu \\ \vecbf{\kappa} \end{pmatrix}\Theta^{\vecbf{\kappa}}(\vecbf{\tilde{d}} - \hat{d}) f_{\vecbf{d}}(\vecbf{\tilde{d}})\de\vecbf{\tilde{d}}\\ &= 
\frac{1}{\numvox^\nu}\sum_{|\vecbf{\kappa}| = \nu} \begin{pmatrix} \nu \\ \vecbf{\kappa} \end{pmatrix} \left[1 - F_{\vecbf{d}_\kappa}(\hat{d}\cdot\vecbf{1}_{\nu})\right]
\end{aligned}\label{eq:app:multiindex_Dvh}
\end{equation}
where $F_{\vecbf{d}_\kappa}(\hat{d}\cdot\vecbf{1}_{\nu})$ corresponds to the evaluation of a $\nu$-variate marginal probability and $\vecbf{1}_\nu \in \mathbb{R}^{\nu}$ is a vector with each of the $\nu$ components equal to $1$. For example, in the case of $\nu = 3$ and $\numvox = 4$, given an index combination $\vecbf{\kappa} = (2, 0, 1, 0)$ (satisfying the sum condition $|\vecbf{\kappa}| = \nu = 3$), the trivariate probability $F_{\vecbf{d}_{1;1;3}}((\hat{d},\hat{d},\hat{d})^\transposed)$ needs to be evaluated. Note that the possible \enquote{doubling} of an index (\ie $\kappa_i > 1$) can also be eliminated in the underlying integral using $\Theta(x)^{\kappa_i} = \Theta(x)$. This reduces the given example to an evaluation of a bivariate probability $F_{\vecbf{d}_{1;3}}((\hat{d},\hat{d})^\transposed) = F_{\vecbf{d}_{1;1;3}}((\hat{d},\hat{d},\hat{d})^\transposed)$.

\end{document}